\documentclass[12pt]{article} \usepackage{amsmath} \usepackage{amssymb}
\usepackage{epsfig}
\usepackage{textcomp}

\newcommand{\alg}{\Sigma}

\newcommand{\cA}{\mathcal{A}}
\newcommand{\cB}{\mathcal{B}}
\newcommand{\cC}{\mathcal{C}}

\newcommand{\cM}{\mathcal{M}}
\newcommand{\cP}{\mathcal{P}}
\newcommand{\cR}{\mathcal{R}}
\newcommand{\cX}{\mathcal{X}}
\newcommand{\cY}{\mathcal{Y}}
\newcommand{\cZ}{\mathcal{Z}}

\newcommand{\Om}{\Omega}

\newcommand{\tOm}{\tilde{\Omega}}
\newcommand{\tgamma}{\tilde{\gamma}}

\newcommand{\tTheta}{\tilde{\Theta}}

\newcommand{\hgamma}{\hat{\gamma}}

\newcommand{\halg}{\hat{\alg}}

\newcommand{\hf}{1/2}

\newcommand{\pow}{\operatorname{pow}}

 at10pt

\begin{document}
\title {Causality, Bell's theorem, and Ontic Definiteness} \author{Joe Henson\footnote{jhenson@perimeterinstitute.ca} } \maketitle
\begin{abstract}
Bell's theorem shows that the reasonable relativistic causal principle known as ``local causality'' is not compatible with the predictions of quantum mechanics.  It is not possible maintain a satisfying causal principle of this type while dropping any of the better-known assumptions of Bell's theorem.  However, another assumption of Bell's theorem is the use of classical logic.  One part of this assumption is the principle of \textit{ontic definiteness}, that is, that it must in principle be possible to assign definite truth values to all propositions treated in the theory.  Once the logical setting is clarified somewhat, it can be seen that rejecting this principle does not in any way undermine the type of causal principle used by Bell.  Without ontic definiteness, the deterministic causal condition known as Einstein Locality succeeds in banning superluminal influence (including signalling) whilst allowing correlations that violate Bell's inequalities.  Objections to altering logic, and the consequences for operational and realistic viewpoints, are also addressed.
\end{abstract}
\vskip 1cm
%=====================================================

\section{Introduction}

Bell's theorem \cite{Bell:1964,Bell:1987,Bell:1990} encapsulates by far the most mysterious aspect of quantum mechanics, often summed up as ``quantum non-locality''.  Indeed, Spekkens has put forward an ingenious, but distinctly unmysterious, hidden-variable model that qualitatively reproduces practically every interesting feature of quantum mechanics (interference, no-cloning, \textit{etc.}) \textit{apart} from nonlocality \cite{Spekkens:2004a}.  The theorem demonstrates that a reasonable causal principle which Bell named ``local causality'' is incompatible with the predictions of quantum mechanics, given some other seemingly innocuous assumptions, chiefly ``freedom of settings'' from correlations with events in the past that are relevant to the experiment, and of course the use of standard logic.  As has been noted by several authors, the assumption of local causality can be broken down into two parts, either of which can be rejected \cite{Butterfield:1992a,Brown:2002a}.  The first is the assumption that there should be a Principle of Common Cause (PCC) \cite{Reichenbach:1956a}, that is, that correlations between events require a causal explanation.  In this view, if event $A$ is correlated to $B$, either the events are directly causally related, or the correlation can be explained by a ``common cause'' $C$ (additionally to this, a cause must be to the past of its effect).  The second is the principle of Relativistic Causal Structure (RCS), which states that the relativistic spatiotemporal causal structure should be used to define the past of any spacetime region, a position forced on us if we accept relativity\footnote{Thus, it is not accurate to say that Bell's theorem shows a contradiction between a realistic world-view and relativity (given the other assumptions); the two can be made compatible by dropping the PCC, just as they are in operational interpretations \cite{Brown:2002a}.}.

This creates a difficult choice for anyone attempting to create a satisfying interpretation for quantum mechanics.  At least one of Bell's assumptions must be dropped.  Approaches such as deBroglie-Bohm pilot wave theory \cite{Bohm:1952a,Bohm:1952b} and those based on Nelson's mechanics \cite{Nelson:1966a,Smolin:2006bw} drop the RCS assumption, whereas the GRW ``spontaneous state-vector collapse'' program \cite{Ghirardi:1986a} in Bell's ``flash'' ontology \cite{Bell:1987ch22}, Kent's ``Real Worlds'' program \cite{Kent:2007gd} and arguably the consistent histories program \cite{Hartle:1992as} (to the extent that it aims at producing a unique probability distribution of over a set of events) all drop the PCC assumption.  Suggestions have been made that the freedom of settings assumption could be dropped instead (either in the guise of past events influencing setting choices, or of ``backwards causation'') \cite{Cramer:1986a,Morgan:2006a}.

These considerations lead us to the first main argument of this paper.  Bell's theorem creates the great conceptual difficulties it does for the following reason:  it seems to preclude the possibility of recovering a satisfying causal principle in any theory that contains quantum mechanics.  Each of the options currently on the table leaves something to be desired in this regard, as discussed at greater length below.  This issue is distinct from the measurement problem.  Leaving the measurement problem unsolved simply begs the question, as we will see, and is, at bottom, a particular way of dropping the PCC as an assumption.  Below, it is maintained that a satisfying causal principle, given what we know of physics, should be relativistic.  This at least means that it should ban superluminal signalling.   As we will see, without any one of the three assumptions mentioned above, this cannot be achieved, and so the lack of superluminal signalling remains unexplained by a causal principle in interpretations which drop one of these -- which is distinctly unsatisfying.  However, a satisfying causal principle needs to be more than simply a ban on superluminal signalling, for reasons that are also addressed.  Local causality, or its deterministic cousin, Einstein locality, are better examples of satisfying causal conditions.

Why should we want to reinstate such a causal principle at all?  The reasons for this are treated at greater length in an appendix to this article.  The most basic reason is that doing so may enhance our understanding of present theory, and increase theoretical progress.  Also, if a relativistic version of the PCC could be made compatible with quantum mechanics, it would be a restrictive condition on theories.  This could serve as a guiding principle in the search for new theories. For example, this idea has already been applied in the context of the causal set approach to quantum gravity, at least in a classical setting.  In this case, a stochastic toy model was developed by taking a general dynamical framework and then selecting a physically appealing set of dynamics by applying basic physical principles, including a causal principle \cite{Rideout:2000a}.

The one path left to us is the alteration of logic.  Curiously, although the use of nonstandard logics has often been considered in the study of the foundations of quantum mechanics, this has never been motivated by the possibility of evading Bell's theorem.  Before the problem of Bell's theorem became apparent, Birkhoff and von Neumann attempted to consistently incorporate all projection operators as propositions in one ``quantum logic'' \cite{Birkhoff:1936a,Dalla:2002}, while Reichenbach's three-valued logic approach \cite{Reichenbach:1944a} sought first and foremost to give meaning to complimentary statements, and to complementarity itself, inside the logic being employed, rather than declaring some statements meaningless (although he did come close to some aspects of the discussion below in his treatment of EPR ``causal anomalies'').  Most recently, topos theory has been used to incorporate contextuality in a pseudo-realist setting \cite{Isham:1998jv,Doring:2007ib}, whist anhomomophic logic was introduced to deal with interference in an observer-independent sum-over-histories account of quantum theory \cite{Sorkin:2006wq,Sorkin:2010kg,Dowker:2007kz}.

Altering logic to evade Bell's theorem seems at first like a very drastic and unnatural step to take, which could only be motivated by desperation to avoid the difficulties of other approaches.  Several questions immediately arise.  Firstly, our causal principle of local causality, and freedom of settings, are built on the foundation of normal logic.  At first glance, then, it seems that we would have to reject local causality with any attempt to alter logic -- a case of throwing out the baby with the bathwater.  Proposing a different causal principle in a different logical setting may, therefore, seem therefore like nothing more than an arbitrary proposal that could have been designed to reach the desired conclusion, rather like moving the goalposts in order to score.   Also, several related worries may arise about the connection to the everyday world of standard logic.  Can causation itself be given a satisfying meaning in any particular logical framework?  And, when reasoning \textit{about} the new logical system, are we not applying classical logic in any case?  The justifications for, and ramifications of, taking so radical a step should be considered carefully.

We will see that these concerns can be answered in at least one case of altered logic.  First, a third ``intermediate'' truth value is added between true and false.  This can be done without changing anything in the standard picture, if we treat the indefinite truth value as signifying ignorance of the ``real'', definite, truth value.  The break with standard logic comes when we question whether there need be a definite truth value for all propositions in all circumstances.  A principle along these lines is dubbed \textit{ontic definiteness} below.  As we will see, rejecting this assumption does \textit{no damage at all} to the definition or meaning of Einstein locality or the freedom of settings; together they still ban superluminal signalling, for instance.  Even so, this move does allow us to evade Bell's theorem.  Dropping ontic definiteness thus allows us to preserve a satisfying causal principle without running into a contradiction with the predictions of quantum mechanics.  Rather than being a desperate measure forced on us by Bell's theorem, it is arguable that this innovation would have been welcomed by many philosophers and mathematicians independently of the problem of Bell's theorem.  Besides this -- whether or not this should affect our judgements -- it is not as far from common sense as many ``quantum logics'' that have been proposed.

The structure of the paper is as follows.  In section \ref{s:bell}, a framework is given in which a relativistic causal principle can be defined, as an example of what is being searched for.  Section \ref{s:assumptions} discusses the consequences of dropping assumptions of Bell's theorem for this type of principle.  In section \ref{s:logic}, the assumption of standard logic is addressed, and the assumption of ontic definiteness is defined, leading to a discussion of how quantum correlations in the EPRB experiment may be given a causal explanation. A discussion of the consequences of this view, and further possible objections, is given in section \ref{s:discussion}.  The appendices deal with justifications for insisting on a satisfying causal principle (appendix \ref{a:causal}), relations between various causal principles (appendix \ref{a:comparing}), and a comparison to Reichenbach's three-valued logic (appendix \ref{a:reichenbach}).

\section{A Causal Principle}
\label{s:bell}

Before explaining anything further about maintaining a satisfying causal principle when faced with Bell's theorem, it is worthwhile to consider why we would want to do so.  Since it is somewhat outside the main topic of the article, comments on this topic are relegated to Appendix \ref{a:causal}.  All that is necessary here is to make the following modest assertion: finding a satisfying causal principle that is compatible with quantum theory may help us to make scientific progress.  The intuition behind this will be clarified in the process of defining a causal principle and examining the issue in the context of Bell's theorem.

The next question is: what exactly is meant by a satisfying causal principle?  We can now turn to a discussion of a good example\footnote{It is worth noting here that, despite some discussion, expressions of the type put forward by Reichenbach and Bell have been convincingly defended from various claims that principles of common cause might reasonably be made significantly weaker \cite{Butterfield:1992a,Butterfield:2007a,Uffink:1999,Henson:2005wb}.  See also the comments at the end of the following section.}.

\subsection{Einstein locality}
\label{s:el}

Let us first consider a deterministic, relativistic principle of causality, Einstein Locality (a term apparently introduced by Bell and d'Espagnat \cite{Selleri:1978}).  Discussion will be limited, for the remainder of this article, to theories with fixed background spacetime.  EL is a restriction on non-probabilistic theories, which we can define in the following way.  First, we consider a space of histories $\Omega$.  Each history is supposed to represent a full specification of all physical properties of the system being modelled at all times.  Events are subsets of $\Omega$. The word ``event'' here is taken to mean something that may or may not occur according to the theory, \textit{i.e.}~events correspond to propositions in the theory, and the subset of $\Omega$ is the set of histories for which the proposition is true.  There is a Boolean algebra of events $\alg$, which contains all the events that we can ask about in this theory\footnote{For the purposes of this article it will be convenient to ignore the complications that arise for history spaces with uncountable cardinality, in which case the algebra should more properly be a Boolean sigma-algebra.}.  This simply means that for questions about pairs of events, logical operations are defined in the normal way.  Each history $\gamma \in \Omega$ corresponds to a truth valuation on this algebra, technically a homomorphism $\gamma: \alg \rightarrow \{0,1\}$, where $\{0,1\}$ is the two element Boolean algebra.  Here, $0$ is taken to mean ``false'' and 1 ``true''.  For instance, event $A$ might signify that there is a ball in a particular box, and if $\gamma(A)=1$, that would be true for the history $\gamma$.  That $\gamma(\, \cdot)$ is a homomorphism means that the usual rules of logic apply to the truth values, for instance, if ``$A$'' is false then ``not $A$'' must be true.  We will use set theoretic notation so that $A \cap B$ means, in words, ``$A$ and $B$'', $A^c$ (the set complement) means ``not $A$'', and so on.

We also need to introduce the background spacetime, which we call $\cM$.  For our purposes this $\cM$ can be defined as a weakly causal Lorentzian manifold\footnote{This is a differentiable manifold equipped with a Lorentzian metric that admits no closed causal curves, allowing a well-defined causal relation between all pairs of points.} (although the formalism could be just as easily applied to a discrete causal set).  To say that our spacetime is equipped with a spatiotemporal causal structure means that there is a partial order $\prec$ on its points, where $x \prec y$ signifies that point $x$ is to the past of point $y$.  We call subsets of the spacetime ``regions''.  For a region $\cR \subset \cM$, the ``causal past'' $J^-(\cR)$ is defined to be all those points that are to the past of (or equal to) any point in $\cR$.

Each event occurs somewhere in our fixed background spacetime.  Thus, we define a map between regions and subalgebras of $\alg$, $\Delta: \pow(\cM) \rightarrow \pow(\Sigma)$, where $\pow(\cM)$ is the set of all subsets of $\cM$ and $\pow(\Sigma)$ is the set of all subalgebras of $\Sigma$.  That is, to each region $\cR$ we assign a a Boolean subalgebra of $\alg$, which we label $\alg|_\cR$, and we say that an event $A \in \alg|_\cR$ is \textit{associated to} the region $\cR$.  This amounts to saying that, if a set of events is associated to a region, then logical combinations of those events should also be associated to that region.  These ``regions of association'' must obey certain consistency conditions which we need not go into in detail here (the reader is referred to \cite{Henson:2005wb}).  It will suffice to say that if $\cR \subset \cR'$ then $\alg|_\cR$ is a subalgebra of $\alg|_{\cR'}$.  Truth valuations can be restricted to any subalgebra in the obvious way, and we will denote the restriction of $\gamma(\, \cdot)$ to the subalgebra associated to $\cR$ as $\gamma|_{\cR}: \alg|_\cR \rightarrow \{0,1\}$.  The collection of all such restrictions $\gamma|_{\cR}$ of truth valuations corresponding to histories $\gamma \in \Omega$ will be called $\Omega|_{\cR}$.

A non-probabilistic theory on the history space also provides a subset $\Theta \subset \Omega$ that contains the histories allowed by the theory.  For instance, if $\Omega$ was the set of scalar field configurations $\phi(x)$, $\Theta$ might contain only the solutions to a particular field equation.  We can imagine the theory as a pack of cards, each representing a history in $\Theta$.  If, by experiment and observation, we know the truth values of a certain set of events, we can go through the deck, keeping the cards that agree with our knowledge and throwing the rest away (events whose truth values we don't know are allowed to be true or false on the cards that we keep).  If we have no cards left, our theory has been falsified.  Otherwise, we can use the remaining cards to make inferences about other events (including predictions), if they have the same truth value on all of those cards.

%Similarly to above, the collection of all $\gamma|_{\cR}$ such that $\gamma \in \Theta$ will be called $\Theta|_{\cR}$.  This represents all of the full specifications of events in $\cR$ allowed by the theory.

We will call two events ``correlated'' (resp. ``anticorrelated'') if and only if they share the same truth value (resp. have opposite truth values) for all histories in $\Theta$.  Another way to state this is that $A$ and $B$ are correlated if and only if $A \cap \Theta=B \cap \Theta$.  We will say that $A$ is \textit{weakly correlated} to $B$ if and only if $A \cap \Theta \subset B \cap \Theta$, so that $\gamma(A)=1$ implies that $\gamma(B)=1$ for all histories in $\Theta$.  Weak correlation includes the case of full correlation.  Finally, if we want to discuss probabilities for events, we can put a probability distribution on $\Omega$ that only has support in $\Theta$.

\begin{figure}[ht]
\centering \resizebox{4.5in}{1.5in}{\includegraphics{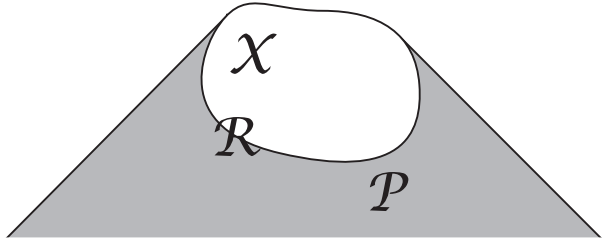}}
\caption{\small{
A ``spacetime diagram'' of the regions involved in the definition of Einstein Locality (here $\cM$ is represented as 2D Minkowski space in a usual spacetime diagram, with time running up the page).  The region $\cR$ is the union of the unshaded region $\cX$ and the shaded past region $\cP=J^-(\cX) \backslash \cX$ within (and including) the light-cone.  According to EL, the truth values of events in $\cP$ must determine the truth values of all events in $\cR$.
\label{f:el}}}
\end{figure}

We now have enough machinery in place to define Einstein Locality.  Consider a region $\cX$, its past $\cR=J^-(\cX)$ and its ``exclusive past'' $\cP=J^-(\cX) \backslash \cX$ where $\backslash$ is set difference, as shown in figure \ref{f:el}.  The idea is that what happens in $\cR$ is determined by what happens in $\cP$\footnote{The reason that we use the region $\cR$ rather than $\cX$ is to make our condition sound even if we allow that some events associated with $\cR$ may not be logical combinations of events associated with $\cX$ and events associated with $\cP$.  This type of condition is known as ``separability of events'', and is called property (iv) in \cite{Henson:2005wb}.}.  So, if we fix the truth value of every event that occurs in $\cP$, then this should fix all truth values in $\cR$.

\paragraph{Einstein locality:} Consider a nonprobabilistic theory defined by $\{\Omega,\alg,\Theta,\cM,\Delta \}$, and regions $\cX$, $\cR=J^-(\cX)$ and $\cP=J^-(\cX) \backslash \cX$. The theory is \textit{Einstein Local} if the following condition holds for all regions $\cX$:

$\forall \gamma_\cP^* \in \Omega|_\cP$, $\exists \gamma_\cR^* \in \Omega|_\cR$ such that $\forall \gamma \in \Theta$, if $\gamma|_\cP=\gamma_\cP^*$ then $\gamma|_\cR=\gamma_\cR^*$.

\paragraph{} That is, if only we knew everything about events in $\cP$, we could infer the truth value of all events in $\cR$.  That is what is meant by events in $\cP$ determining events in $\cR$.  To explain this principle let us return to the analogy of the deck of cards.  We wish to keep in our hand only those cards that are possible histories.  Einstein Locality says that, if we condition on all events associated with the past region $\cP$, so that we only hold cards with a particular set of truths values for these (called $\gamma_\cP^*$), then we will find that, for all events in $\cR$, the same truth values (given by $\gamma_\cR^*$) will be given on every remaining card.

There is another way of stating Einstein Locality:

\paragraph{Einstein Locality (second formulation):} Consider a nonprobabilistic theory defined by $\{\Omega,\alg,\Theta,\cM,\Delta \}$, and regions $\cX$, $\cR=J^-(\cX)$ and $\cP=J^-(\cX) \backslash \cX$. The theory is \textit{Einstein Local} if the following condition holds for all regions $\cX$:

$\forall A \in \alg|_\cR$, $\exists B \in \alg|_\cP$ such that $\gamma(A)=\gamma(B)$ $\forall \gamma \in \Theta$.

\paragraph{} In words, this means that for all events associated to $\cR$ there is an event associated to $\cP$ that is correlated to (or ``determines'') it.  The determining past event $B$ will sometimes be termed the \textit{causal antecedent} of $A$ below.  The proof that these two definitions are equivalent is relegated to appendix \ref{a:twodefs}.

\begin{figure}[ht]
\centering \resizebox{4.5in}{1.5in}{\includegraphics{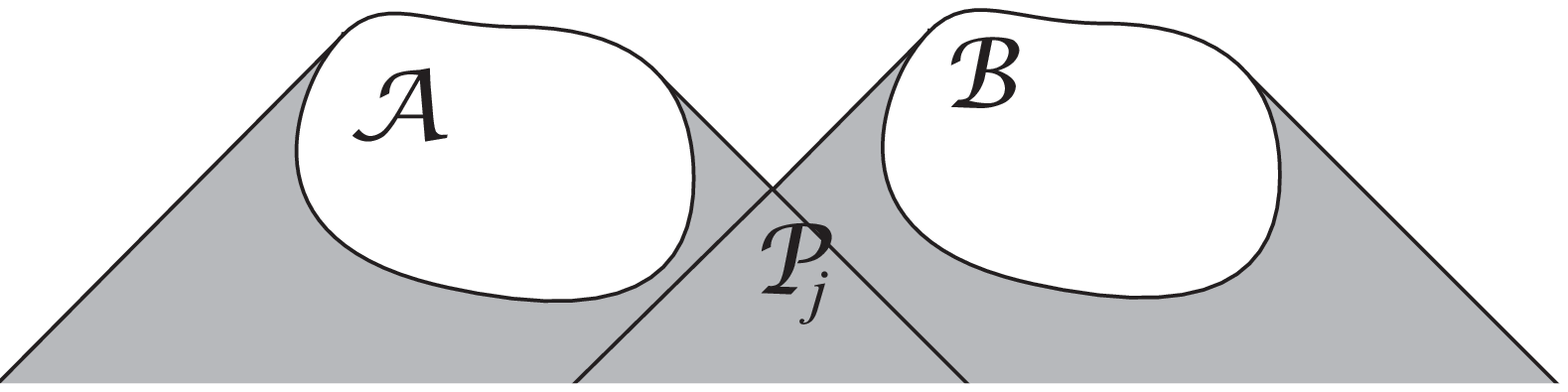}}
\caption{\small{
A ``spacetime diagram'' of the regions involved in the definition of the Nonprobabilistic Principle of Common Cause (joint past version).  The spacelike regions $\cA$ and $\cB$ are shown unshaded, while $\cP_j=J^-(\cA) \cup J^-(\cB) \backslash (\cA \cup \cB)$ is the entire shaed region between (and including) both lightcones.  According this PCC, any correlated pairs of events in regions $\cA$ and $\cB$ must be correlated to a further event in $\cP_j$.
\label{f:pcc}}}
\end{figure}

This condition has a close link to Reichenbachian principles of common cause \cite{Reichenbach:1956a,Henson:2005wb}.  Einstein locality encapsulates a nonprobabilistic PCC (NPCC).  The following is a ``joint past'' version of this principle:

\paragraph{NPCCj:} Consider a nonprobabilistic theory defined by $\{\Omega,\alg,\Theta,\cM,\Delta \}$, two spacelike regions $\cA$ and $\cB$, and their ``joint past'' $\cP_j=J^-(\cA) \cup J^-(\cB) \backslash (\cA \cup \cB)$ as shown in figure \ref{f:pcc}. The theory obeys the NPCCj if the following principle holds: for any pair of events $A \in \alg|_\cA$ and $B \in \alg|_\cB$, if $A$ and $B$ are correlated, then there exists an event $C \in \alg|_{\cP_j}$ that is correlated to both $B$ and $C$.

\paragraph{} This follows from the second formulation of Einstein locality: that principle requires that event there is an event $C$ in the past (that is, associated to $J^-(\cA)$, which is a subset of the joint past $\cP_j$) such that $\gamma(A)=\gamma(C)$ $\forall \gamma \in \Theta$, and if $\gamma(A)=\gamma(B)$ $\forall \gamma \in \Theta$, then $C$ is also correlated to $B$.  By considering the complement of events we can easily extend this reasoning to anticorrelated events as well.  There is also a ``mutual past" version of this principle:

\paragraph{NPCCm:} Consider a nonprobabilistic theory defined by $\{\Omega,\alg,\Theta,\cM,\Delta \}$, two spacelike regions $\cA$ and $\cB$, and their ``mutual past'' $\cP_m=J^-(\cA) \cap J^-(\cB)$ as shown in figure \ref{f:pcc}. The theory obeys NPCCm if the following principle holds: for any pair of events $A \in \alg|_\cA$ and $B \in \alg|_\cB$, if $A$ and $B$ are correlated, then there exists an event $C \in \alg|_{\cP_m}$ that is correlated to both $B$ and $C$.

\paragraph{} These two versions on the NPCC are in fact equivalent, as is proven in appendix \ref{a:twoPCCs}.  Thus, both of them follow from Einstein Locality.

What is the relation between Einstein Locality and other causal principles?  In the case of a non-relativistic theory, instead of a relativistic spatiotemporal causal structure, we simply put all events that are associated to spacetime regions before time $t$ to the past of all events that are associated to regions after or at that time.  Above, substituting the relativistic causal structure of our Lorentzian manifold $\cM$ with this kind of causal structure would give the kind of determinism found in Newtonian theories (sometimes called Millsian causation \cite{Norton:2003a}).

One might wonder if it is necessary to have strict determinism to have a good causal principle that incorporates the PCC and bans superluminal influence.  Indeed, for this purpose is not necessary to have everything in $\cR$ determined by events in $\cP$, but only necessary that, once we have conditioned on all events in $\cP$, nothing more about the truth values of event in $\cR$ can be determined by conditioning on any events that occur spacelike to $\cR$.  This line of argument leads to a weaker condition.  However (and although giving a thorough defence of this would require a separate article), I claim that (a) this weakening does not evade Bell's theorem and (b) we will be able to maintain the stronger condition of Einstein locality when we alter logic.  Because of this, the weaker condition is not relevant for the following discussion.

Beyond this, there are the probabilistic formulations such as Bell's local causality.  It is not hard to convince oneself that these follow from Einstein locality in the following sense.  If we take the histories in $\Theta$ and put a probability distribution over them, then, if the original theory was Einstein local, the resulting probabilities theory will obey a version of local causality called SO2 in \cite{Henson:2005wb}.  This derivation is left for appendix \ref{a:ELandSO2}.  Again, it is well known that using the probabilistic version does not evade Bell's theorem.  There are several other formulations of probabilistic, relativistic causal principles, but in previous work it was conjectured that all of these are either problematic, equivalent to local causality, or incorporate additional inessential assumptions \cite{Henson:2005wb}, and steps have been taken towards showing this (see also \cite{Butterfield:2007a}).  The conclusion, again, is that Einstein locality is a reasonable condition to concentrate on for the purposes outlined in the introduction of this article.  The important thing to keep in mind is that Einstein locality expresses the principle that there are no superluminal influences, as all events are determined by events in their past light-cone.  It encompasses a combination of a principle of common cause, and relativistic causal structure.  In particular, when the theory describes agents and devices, Einstein locality bans superluminal signalling.

\subsection{Einstein locality and superluminal signalling}
\label{s:signalling}

As explained in appendix \ref{a:ban}, a distinction needs to be drawn between a simple operational ban on superluminal \textit{signalling}, and causal conditions like EL which ban superluminal \textit{influence}.  Nonetheless, since signalling is a type of influence, any relativistic causal condition worth its salt should imply a ban on superluminal signalling.  Fortunately it is not hard to see that EL does this.

We can now give a straightforward account of how Einstein locality gives rise to a ban on superluminal signalling.  Consider an agent in a region $\cA$ trying to influence something in a spacelike region $\cB$.  Some event $A \in \alg|_\cA$ is to be regarded as the input event, and some event $B \in \alg|_\cB$ will be the output.  In the context of non-probabilistic theories set out above, we can distinguish \textit{strong} and \textit{weak} signalling.  Let us consider strong signals first. A strong signal is successful if and only if $A$ and $B$ are correlated, which as before means that $\gamma(A)=\gamma(B)$ $\forall \, \gamma \in \Theta$.  This is not sufficient to define signalling, however.  For example, if two newspapers delivered to two different houses have the same headline, this does not constitute a signal between the newspapers.  Both events were determined by a common cause in the past.  Thus we must impose some condition to make sure that event $A$ is a ``free setting'': that is, $A$ is free from the influence of past events that may also influence $B$.  One requirement is that, in our model, $A$ is not determined by any event in its past region $J^-(\cA) \backslash \cA$.  If we are applying EL, we must in this case exempt the event $A$ from being determined, basically because we have left out its causal antecedents from our model.  For instance, we might decide $A$ based on what we had for breakfast. ``Freedom of settings'' means assuming that what we had for breakfast, or any other such factor, is irrelevant when modelling the system in question.  The event $B$ however is not free.  With $A$ ``free'' in this sense, if $\gamma(A)=\gamma(B)$ $\forall \, \gamma \in \Theta$, then we must admit that there is superluminal signalling from $\cA$ to $\cB$.

It can now be seen that this can be ruled out by applying EL.  Consider the region $\cZ=J^-(\cB) \backslash J^-(\cA)$.  Because $\cB \subset \cZ$, it follows that $B \in \Sigma|_\cZ$, and so by Einstein Locality $B$ requires a causal antecedent in $J^-(\cZ)=J^-(\cA) \cap J^-(\cB)$, that is, an event $C$ such that $\gamma(B)=\gamma(C)$ $\forall \, \gamma \in \Theta$.  But if $\gamma(A)=\gamma(B)$ $\forall \, \gamma \in \Theta$ this implies $\gamma(A)=\gamma(C)$ $\forall \, \gamma \in \Theta$ which contradicts our freedom of settings assumption for $A$, because $C$ is associated to $J^-(\cA) \cap J^-(\cB)$ and thus is associated to the past of $\cA$ \footnote{This conclusion might seem surprising: why cannot the chain of causal antecedents of $B$ remain outside of the past of $\cA$ arbitrarily far into the past, or until some initial cosmological hypersurface?  Such a scenario would imply that the setting event was correlated to some initial cosmological event outside its past light-cone, or events arbitrarily far in the past, however, and it is reasonable that freedom of settings and EL together ban this.  A similar argument is dealt with in appendix \ref{a:twoPCCs} (see \cite{Henson:2005wb} for more on this point).}.

Weak signalling means that the truth value of the signal \textit{can} affect the value of the output, but does not in all circumstances.  There is a weak signal between $A$ and $B$ if and only if $A$ is a free setting, and $B$ is weakly correlated to $A$.  In this case, $\gamma(B)=0$ whenever $\gamma(A)=0$ for all histories in $\Theta$, and so the truth value of $A$ can determine the truth value of $B$, although the events may not be fully correlated in the sense already defined.  Changing the setting could thus affect the truth value of the event $B$.  The definition could just as easily have been that $A$ is weakly correlated to $B$ (rather than $B$ to $A$), but this is a matter of convention if $A^c$ is also regarded as a signal, which it will be.

This line of thought also shows that our condition for an event like $A$ to be a free setting is incomplete: we do not want any event $C$ in $J^-(\cA) \backslash \cA$ to be even weakly correlated to $A$.  If it were, $\gamma(A)=1$ is forced on us if we have $\gamma(C)=1$, and so $A$ can hardly be described as free from influence from its past.  Again, Einstein Locality ensures that there can be no superluminal signalling, even of this weak variety.  We previously established that, if EL holds, there is an event $C$ associated to the past of $\cA$ that is correlated to $B$. If $B$ is weakly correlated to $A$, then $C$ must also be weakly correlated to $A$, violating our strengthened freedom of settings condition.

In the probabilistic setting, freedom of settings again means that the setting $A$ should not be probabilistically correlated to any event in the past region (and neither should it become so after we condition on any other event in the past).  Here we are using the word correlation in the usual probabilistic sense. This is the assumption that goes into Bell's theorem.

\section{Dropping assumptions of Bell's theorem}
\label{s:assumptions}

Having discussed an example of a satisfying causal principle, we now examine the effects on the maintenance of such a principle if we drop assumptions of Bell's theorem.

\subsection{Dropping counterfactual definiteness?}

It has been claimed by some that, as well as local causality, Bell's theorem depends on a certain ``hidden-variables'' assumption.  Thus those who object to the lack of local causality in Quantum  Mechanics suggested by Bell's theorem, it it argued, need only reject the idea of hidden variables.  This view is summed up by Mermin \cite{Mermin:1993} when he says ``to those for whom nonlocality is anathema, Bell's Theorem finally spells the death of the hidden-variables program.''

For simplicity let us consider measurements with outcome valued in $\{0,1\}$ (\textit{i.e.}~measurements corresponding to single events).  Counterfactual definiteness asserts that, for every possible measurement, there must be an event (occurring in the same region as the measurement) that shares the truth value that \textit{would have} been obtained if that measurement was performed, irrespective of whether the measurement actually was performed.  This is sometimes cited as an assumption of Bell's theorem, leading to conclusions like Mermin's.   One interpretation of the sentence quoted above is that there is a choice between locality and counterfactual definiteness, and that, in operational interpretations in particular, we can have a satisfactory resolution to the problem of Bell's theorem if we drop counterfactual definiteness.

This interpretation is mistaken, however.  Norsen (quoting other earlier treatments) convincingly argues that counterfactual definiteness is \textit{not} a separate assumption, but in fact follows from local causality (and the results of quantum mechanics which specify that perfect correlations between some outcome events can be achieved in the EPRB set-up) \cite{Norsen:2006a}.  Indeed, it is clear in Bell's last published account of his result \cite{Bell:1990} that the argument begins with causality and leads to counterfactual definiteness.  Thus, whether or not we start with a bias on the subject of hidden variables, if we accept Bell's local causality as a good formalisation of the idea of a relativistic causal principle, then we are forced to accept counterfactual definiteness.  In other words, when standard logic is assumed, rejecting counterfactual definiteness amounts to rejecting local causality (and Einstein Locality).

The lack of superluminal signalling is often mentioned in the operational context, to justify quantum theory against the principle of relativity.  Be that as it may, lack of superluminal signalling cannot be substituted for local causality; it is not in itself a good relativistic causal principle of the type being searched for here, as argued in appendix \ref{a:ban}.  The problem of quantum nonlocality still remains.  Thus, ``to those for whom nonlocality is anathema'', \textit{all} of the options presently on the table should be equally unsatisfying.

The conclusion is that there is no separate assumption of Bell's theorem here.  Despite this, later we will see that there is a way to resurrect the counterfactual definiteness ``loophole'' while maintaining Einstein Locality.  This is made possible by dropping another, deeper assumption.

\subsection{Dropping the PCC}
\label{s:droppcc}

Dropping the principle of common cause itself clearly amounts to throwing in the towel as far as maintaining a causal principle is concerned. In this approach, there are still prospects for doing away with measurement as a fundamental notion in quantum mechanics, but one gives up on another goal of quantum interpretation: that of inferring facts about the quantum world using a causal principle, as briefly argued in appendix \ref{a:causal}.  In interpretations like Kent's Real Worlds \cite{Kent:2007gd}, and consistent histories \cite{Hartle:1992as}(in one reading), the goal is to extract from the quantum formalism some probability distribution on a coarse-grained algebra of events, which should at least be larger than any set of questions that we could ever answer operationally.  But confronted with the strange correlations of the EPRB experiment, one simply claims that correlations do not need to be explained.  Although the production of the original Bell state in the past of both wings of the experiment seems to be the ``cause'' in some sense, this has not been formalised.  But how, in this view, do we explain the lack of superluminal signalling in relativistic quantum theories?  It can have nothing to do with a deeper lack of superluminal influence, talk of which we have forsworn; it is merely an unexplained fact.  GRW ``state-vector-collapse'' models \cite{Ghirardi:1986a}, most clearly in Bell's flash ontology \cite{Bell:1987ch22}, have the same feature.

Operational treatments of quantum mechanics should also be put in this category.  In this case the events that we allow in our theory are only those in some set of ``operational'' events, referring explicitly to preparations and observations.  But nothing stops us from conceiving of such theories as probability distributions over this rather impoverished algebra of events (albeit that the set of events is not given by the theory, but defined by fiat, due to the arbitrary split between the classical and quantum worlds), which can as above be located in spacetime regions.  Again, we find that the PCC is not respected; correlations between spacelike events cannot always be explained by conditioning on past events.  As Norsen has noted, then, there is something wrong with the assertion that the assumptions of Bell's theorem can be summed up as ``local realism'', as if there is a separate assumption of realism involved, the rejection of which relieves us of the conundurum of Bell's theorem \cite{Norsen:2007a}.  Rather, operationalist accounts also have to drop the assumption of the PCC, and thus give just as bad an account of causality as some of the so-called ``realist'' alternatives.

I will also claim that many-worlds interpretations should be placed in this category and thus fail to maintain a satisfying causal principle.  The point has been made by Timpson and Brown \cite{Brown:2002a} that the PCC has to be dropped in this case for consistency with relativity\footnote{However,the following discussion in \cite{Brown:2002a} attempts to give a causal account of the EPRB experiment in the many-worlds setting in a different way.  These ideas seem to me inadequate to recover a satisfying causal principle in the sense that it is being searched for here, however.  Discussion of these subtleties will be left for future work.}.

\subsection{Dropping RCS}

Instead we can insist on the PCC, and freedom of settings, but reject relativistic causal structure.  In doing so we can at least maintain a non-relativistic, ``Millsian'' version of causation as in Newtonian theories.  This is the position taken in deBroglie-Bohm pilot wave theories \cite{Bohm:1952a,Bohm:1952b}, or in reworkings \cite{Smolin:2006bw} of Nelson's ``stochastic mechanics'' programme \cite{Nelson:1966a}, for example.  In this view, the lack of superluminal signalling is merely a peculiarity of the theory, which is not related to any objective lack of superluminal influence.  Likewise the apparent relativistic nature of events at the classical level is just an emergent peculiarity that is not reflected in the fundamental theory.

The following discussion will be based on the position that this does not add up to a satisfying causal principle.  We have added extra events to quantum mechanics in order to satisfy the PCC.  But we do not find an ontology that agrees with important physical principles like those of relativity.  Neither do we recover a causal principle that explains the lack of superluminal signalling.  Instead, we end up with experimentally inaccessible events that violate relativity and allow superluminal influences.  It is a reasonable (though not inescapable) conclusion that something has gone wrong: ``to those for whom nonlocality is anathema'', these hypothetical new events provide an unlikely explanation of apparent events, and so are not telling us anything useful.  The new underlying variables seem to be a ``fifth wheel'' in the theory.

\subsection{Dropping Freedom of Settings}

This leaves the possibility of dropping the ``freedom of settings'' assumption.  This seems to provide us with a way to keep both a causal principle and relativity in the face of Bell's theorem.  Given the ``apparent'' events (settings and outcomes) described by Bell, we can ``fine-grain'' the description, adding hypothetical events to the past in order to find a model that contains the apparent events, but which also obeys local causality.  The problem with this approach is that \textit{any} probability distribution over events happening in spacetime can be ``causally completed'' in this way.  This follows from a theorem of Placek's \cite{Placek:2000}\footnote{The theorem does not consider the issue of \textit{where} in spacetime events occur in the same way as has been done in the present work, but this issue can easily be dealt with by putting the added ``hypothetical'' events into the past of all of the original ``apparent'' events.}.  This includes even the most blatant cases of superluminal signalling.  For example, as in the discussion of section \ref{s:signalling}, our agent could be controlling a ``setting'' event $A$ in a region $\cA$ spacelike to region $\cB$, containing event $B$.  Let us say that the events are perfectly correlated.  Without the freedom of settings requirement, we can make a causal model of this simply by adding an event $C$ in the past that is perfectly correlated with $A$ and $B$.  Perhaps $C$ could describe some event involving the decay of a ``conspiriton'' particle, one half of which travels off to determine $B$ whilst the other half goes off to determine the agent's decision on what value of $A$ to choose -- and also anything correlated to her decision, for instance what she had for breakfast, if she chooses the setting on that basis.

Without any further restriction, the proposed relativistic causal principle is completely empty.  In particular, it allows superluminal signalling.  In this case it is not deserving of the name; it is not a satisfying relativistic causal principle.  It is possible that, if we first imposed other conditions on our theory, local causality could regain some power, but then the problem would be justifying calling this set of conditions a causal principle.  As well as banning superluminal signalling, a causal principle should be justifiable on counterfactual grounds.  Although none of these considerations rule out making an interpretation of quantum mechanics based on this loophole, they do critically undermine motivation for doing so rather than pursuing other methods\footnote{Two other ``loopholes'' should be mentioned at this point.  The first is that commented on by Morgan: that Bell correlations could be explained if we allow ``initial correlations'', i.e. correlations between event on some initial spacelike surface before the experiment, or correlations that somehow propagate in from past infinity \cite{Morgan:2006a}.  This is technically distinct from freedom of settings, but the arguments against exploiting it are very similar.  Secondly there is ``backwards causation'', as exploited in Cramer's transactional interpretation \cite{Cramer:1986a}, which can be seen as a different gloss on the rejection of freedom of settings from correlations with past events, and fails to maintain a satisfying causal principle for the same reason.}.

As we have seen, whether or not to solve the measurement problem is only one of the major decisions that must be made when interpreting quantum mechanics.  A second, perhaps equally important choice, is which assumption of Bell's theorem to sacrifice.  The most well-known interpretations all choose to drop either the PCC or relativistic causal structure. Explicitly or implicitly, this choice depends on whether or not we feel that the principle of common cause is so important to physics that it must be preserved, even at the expense of relativity.  We have also seen that dropping any of the three assumptions that are most usually discussed leads to an unsatisfying situation regarding relativistic causal principles. The question now is, can we do better by exploiting the forth assumption, that of standard logic?

\section{A modest alteration in logic}
\label{s:logic}

\subsection{Possible objections}
\label{s:objections}

There are (at least) four objections that may cause some reticence to alter logic when faced with the problem of Bell's theorem.  Objection (1), ``throwing the baby out with the bathwater'', is that all the other assumptions of Bell's theorem are premised on standard logic, and so it would seem that rejecting logic does not allow us to save them.  In answer to this, we will see below that one particular part of normal logic can be removed without doing any damage to a particular formulation of Einstein locality. Objection (2), ``moving the goalposts'', is similar, but directed at the \textit{motivation} for the causal principle after logic has been altered.  The argument is that the whole idea of causality depends on standard logic, and anything that might be called a causal principle in a non-standard logical setting is in fact an arbitrary invention.  To combat this, in the details below it will be shown that such a principle can indeed be satisfying in the sense of banning superluminal influences (including signalling) and having a counterfactual motivation that is practically identical to that of Einstein Locality in the case of standard logic, giving it every bit as much meaning.  Objection (3), ``common sense", is that logic should not be altered on empirical grounds (for debate on this point see \cite{Putnam:1968,Dummet:1976a}); since it applies to the way we reason at all times -- even while considering altering logic -- and to everyday events, standard logic simply cannot be rejected.  Hopefully, the modest alteration prosed below will not draw so much criticism in this regard, for the following reasons.  Firstly, similar things have been proposed outside of the present context, and secondly, the usual reasoning for everyday or scientific purposes is not threatened.  Objection (4), ``the meta-level'' objection, is that even if we had a physical theory based on non-standard logic, we would still have to reason \textit{about} it using standard logic.  Would it not always be possible to reduce the theory to one based on standard logic, and in that case, are we not stuck with the choice between the original three assumptions of Bell's theorem? The details of this objection, and the answer to it, are somewhat more involved, and so discussion of this point will be deferred until we have an example to work with.

Standard logic (which we are applying to events in spacetime) consists of the Boolean algebra of events $\alg$, the standard space of truth valuations $\{0,1\}$ and the standard type of truth valuations or ``histories'' $\gamma$, which are homomorphisms.  Below, the Boolean algebra of events will be kept, but the space of truth valuations will be replaced with $\{0,\hf,1\}$.  Reichenbach also considered such a move as a response to quantum mechanics \cite{Reichenbach:1944a}; the approach here differs significantly from his, as explained in Appendix \ref{a:reichenbach}. The three truth values will first be introduced in a way that does no injury to standard ideas of logic\footnote{The approach taken here could be called a simple type of ``epistemic logic'' \cite{Hintikka:1962}, and also turns out to have similarities to Rescher and Brandom's ``schematic worlds'' \cite{Brandom:1980a}.}.

\subsection{Introducing the intermediate truth value: ``a theory of local knowables''}
\label{s:thirdvalue}

The Boolean algebra $\alg$ of events has already been introduced, along with the idea of a truth valuation or history $\gamma:\alg\longrightarrow \{0,1\}$, and the history space $\Omega$.  The histories $\gamma$ represent possible realities here; we can talk as if one history is the one that actually exists, independent of our knowledge of the truth values of the various events.  Histories like this can been termed \textit{ontic} \cite{Spekkens:2004a}.  But what if we don't know some of the truth values?  To put it another way, we might know that the ontic history lies in a certain subset of $\Omega$, but have no knowledge about it beyond that.  This gives rise to the idea of an \textit{epistemic} history, which describes the state of knowledge rather than the most detailed possible description of the history.  An easy way to describe an epistemic history is by a subset of $\Omega$.  But we can also describe it as a new kind of truth valuation on the algebra of events $\alg$.  Here we need the extra truth value $\hf$, which we will call ``intermediate'' \footnote{This term is meant to be considered philosophically neutral, rather than implying an ``ontologically distinct'' third truth value (which we will term ``indefinite'') or a merely epistemic notion (which we will term ``unknown'').}.  ``Three-valued truth valuations'' will be truth valuations using the three values.  For the purpose of this section (that is, to motivate the new logic), the term ``three-valued'' can be replaced with ``epistemic'', and ``intermediate'' with ``unknown'', without disturbing the meaning.  These three-valued truth valuations can be derived from a set of ontic histories in the following way.

The rule for deriving a three-valued truth valuation $\tgamma_X$ from a set of ontic histories $X$ is as follows: if there exists some $t \in \{0,1\}$ for which $\gamma(A)=t$ $\forall \, \gamma \in X$, then $\tgamma_X(A)=t$, otherwise $\tgamma_X(A)=\hf$.  An equivalent way of saying this is that $\tgamma_X(A)=1$ if and only if $A \supset X$, $\tgamma_X(A)=0$ if and only if $A^c \supset X$, and $\tgamma_X(A)=\hf$ otherwise.  This is exactly the intuitive rule: if we know the history is in some set for which our proposition is always true (resp. false), then the proposition must be true (resp. false); if, on the other hand, all we know is that the history is in some set for which our proposition can be true \textit{or} false, then we don't know its actual truth value.  We will say that $\tgamma_X$ is the three-valued truth valuation derived from the set $X$.  Note that $X$ is itself an event, and the events that are true for $\tgamma_X$ are those that are logically implied by $X$.  Note also that the first formulation of ``deriving a three-valued truth valuation'' given here can also be applied to a set of three-valued truth valuations, rather than ontic ones.  This can be used to further coarse-grain the state of knowledge.

\begin{figure}[ht]
\centering \resizebox{4.5in}{1.5in}{\includegraphics{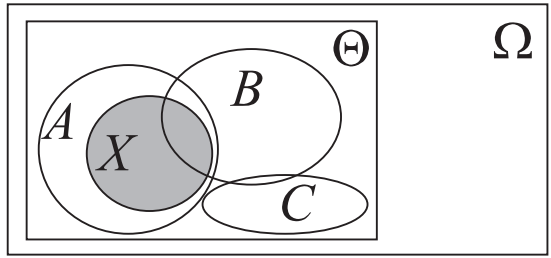}}
\caption{\small{
A Venn diagram depicting a history space $\Omega$.  Each point in the space represents an ``ontic'' history $\gamma \in \Omega$, and the subset $\Theta$ represents all histories allowed by our theory, as explained in section \ref{s:el}.  A three-valued truth valuation $\tgamma_X$ can be derived from the set $X$ as explained in the text.  It can be imagined that it is only known that the ``real'' history is in the set $X$.  In the case shown, $\tgamma_X(A)=1$ because $\gamma(A)=1$ for all $\gamma \in X$.  In contrast, $\tgamma_X(B)=\hf$ and $\tgamma_X(C)=0$.
\label{f:omega}}}
\end{figure}

What is the logic that applies to this new kind of truth valuation?  For definite truth values (meaning 0 and 1), standard truth tables apply.  For intermediate values, there is not always a unique output for a logical operation.  For instance, consider the operation ``and''.  If $\tgamma(A)=\tgamma(B)=\hf$, then $\tgamma(A \cap B)$ can equal $\hf$ or $0$.  It cannot equal $1$ as this would imply that both $A$ and $B$ must have the definite value $1$.  It is possible to set $\tgamma(A \cap B)=0$ however since this does not fix the ontic truth value of either $A$ or $B$.  This logic will not be examined in detail here; it is a sufficient definition that all three-valued truth valuations can be derived from a subset of some ontic history space $\Omega$ in the above sense.

What is a theory now?  In the standard case, it was described by $\{\Omega,\alg,\Theta,\cM,\Delta \}$.  Each member of $\Omega$ is a history $\gamma$ with an associated two-valued truth valuation $\gamma(\, \cdot)$.  In the three-valued case this suggests $\{\tOm,\alg,\tTheta,\cM,\Delta \}$, where $\tOm$ is the space of all subsets of $\Omega$, each of which has an associated three-valued truth valuation, and $\tTheta$ is the subset of these that are also subsets of $\Theta$.  That is, if our three-valued theory is merely to be the epistemic version of the theory $\{\Omega,\alg,\Theta,\cM,\Delta \}$, then $\tTheta$ should just be the set of all three-valued truth valuations that can be derived from subsets of $\Theta$.  In this case we will say that $\tTheta$ has been derived from $\Theta$.  When $\tTheta$ is derived in this way, our ontic theory tells us that the ``real'' history is in the set $\Theta$, and the histories allowed in our three-valued theory just express ignorance about which one it is.  Epistemic theories $\{\tOm,\alg,\tTheta,\cM,\Delta \}$ derived in this way from ontic theories therefore obey the following principle\footnote{It is tempting to simply name this condition ``realism''.  This would chime somewhat with the usage in philosophy, especially the work of Dummett \cite{Dummett:1963a}.  However, this word has long been used in discussions of quantum mechanics to mean something along the lines of ``the opposite of operationalism'', which is certainly not intended here.  Indeed, one of the main points to be made here is that the rejection of this condition need not dismay physicists who fall into the realist camp.}:

\paragraph{Ontic definiteness:} A three-valued theory $\{\tOm,\alg,\tTheta,\cM,\Delta \}$ obeys the principle of ontic definiteness if and only if there exists some subset $\Theta$ of the ontic history space $\Omega$ from which $\tOm$ is derived, such that $\Theta \subset \tTheta$ and all of the indefinite truth valuations in $\tTheta$ can be derived from subsets of $\Theta$.

\paragraph{} In this way we can derive a three-valued theory from any ontic theory, simply by allowing for the truth values of some events to be unknown.  In the card analogy, we now have a larger deck of cards, each one of which represents a truth valuation in $\tTheta$.  If we know the truth values of some events, we can again take all the cards that agree with them and throw the rest away.  As before we can use the remaining cards to make predictions and draw inferences.  In this case, there will be a unique card in our remaining hand, the ``actual epistemic truth valuation'', on which all events with unknown truth values have truth value $\hf$.  But also, there will be many other cards representing other three-valued truth valuations which could become applicable when we come to know more. Amongst these, because of ontic definiteness, our hand will always include at least one of our original set of cards, the ontic histories.

\subsection{Einstein locality revisited}
\label{s:ELrevisited}

What becomes of our causal condition in this setting?  If our original ontic theory $\{\Omega,\alg,\Theta,\cM,\Delta \}$ obeys Einstein Locality, does this impose a condition on the three-valued theory $\{\tOm,\alg,\tTheta,\cM,\Delta \}$ derived from it?  First let us consider the term ``correlation'' that we defined earlier for definite histories.  There, two events $A$ and $B$ are correlated if and only if $\gamma(A)=\gamma(B) \, \forall \gamma \in \Theta$, which is equivalent to $A \cap \Theta =B \cap \Theta$.  If this is true, what does it mean in terms of the three-valued truth valuations in the derived three-valued theory?  Any set $X \subset \Theta$ is either a subset of $A \cap \Theta$, a subset of $A^c \cap \Theta$, or it intersects both.  If $A \cap \Theta =B \cap \Theta$, in the first case we have $\tgamma_X(A)=\tgamma_X(B)=1$, in the second case we have $\tgamma_X(A)=\tgamma_X(B)=0$ and in the third case we have $\tgamma_X(A)=\tgamma_X(B)=\hf$.  So if $A$ and $B$ are correlated in the ontic theory, it follows that $\tgamma(A)=\tgamma(B) \, \forall \tgamma \in \tTheta$ in the derived three-valued theory.  It therefore makes sense to adopt this as the definition of correlation for three-valued theories.  Note that the definition for the three-valued case seems as immediately reasonable, on its own terms, as the original definition (indeed, it has the same form).

Now consider the following:

\paragraph{Einstein locality (three-valued formulation):} Consider an indefinite theory defined by $\{\tOm,\alg,\tTheta,\cM,\Delta \}$, and regions $\cX$, $\cR=J^-(\cX)$ and $\cP=J^-(\cX) \backslash \cX$. The theory is \textit{Einstein local} if the following condition holds for all regions $\cX$:

$\forall A \in \alg|_\cR$, $\exists B \in \alg|_\cP$ such that $\tgamma(A)=\tgamma(B)$ $\forall \tgamma \in \tTheta$.

\paragraph{} This is almost identical to the second formulation of Einstein locality given in section \ref{s:el}.  It is not hard to see that this condition follows if the theory $\{\tOm,\alg,\tTheta,\cM,\Delta \}$ is derived from an ontic theory obeying EL.  The second formulation of EL states that every event $A$ that occurs in $\cR$ is correlated to some event $B$ that occurs in $\cP$.  If two events are correlated in this sense in the ontic theory, it implies that $\tgamma(A)=\tgamma(B)$ for all $\tgamma \in \tTheta$ as well, as explained above.  This shows that the three-valued formulation of EL follows from the more standard definite ones.  It is also important to note that this formulation reduces to the definite formulation when only the original ontic truth valuations are considered.

This new formulation of EL has appeal independent of its derivation from the definite case.  We have discovered that the meaning of correlation in the three-valued case should be that $\tgamma(A)=\tgamma(B)$ for all $\tgamma \in \tTheta$.  If the truth values in question are in $\{0,1\}$ this reduces to the normal definition, and when $\tgamma(A)=\tgamma(B)=1/2$ the definition is also reasonable: if $A$ is determining $B$ (or if they are both determined by some other event) it would make little sense if one held a definite value and the other did not.  With this in mind, the new version of EL simply states that every event is determined by an event in its past.

We also need to make sure that the ban on superluminal signalling follows from EL, as before.  This is not difficult.  Again, we consider a signal event $A$ associated to a region $\cA$, which has to be a ``free setting'', and the output event $B$ associated to a region $\cB$ which is spacelike to $\cA$.  In this case, we also have to note that both $A$ and $B$ must hold definite values; it would not be a signal if the message was unknown to one or both parties.  Freedom of the settings will still mean that we do not want any event $C$ in $J^-(\cA) \backslash \cA$ (at least any such event that we need to include in our model) to be even weakly correlated to $A$ (or $A^c$).  As in the case of correlation, weak correlation also needs a counterpart for the three-valued theory.

In the ontic theory, weak correlation between $C$ and $A$ means that $C \cap \Theta \subset A \cap \Theta$.  Given that the setting $A$ has a definite value, what possible three-truth values can $C$ and $A$ take in a three-valued truth valuation $\tgamma_X(\, \cdot)$?  If $X$ lies completely outside $A$ (\textit{i.e.}~$X \subset A^c$), so that $\tgamma_X(A)=0$, then $\tgamma_X(C)=0$ also.  If on the other hand $X \subset A$, then $X$ can still be completely inside or outside $C$, or neither, so in this case $\tgamma_X(C)$ can hold any of the three truth values.  We find that, when $A$ is definite, weak correlation between $C$ and $A$ should mean that $\tgamma_X(C)=0$ whenever $\tgamma_X(A)=0$ for all $\tgamma \in \tTheta$.  Again, the definition is reasonable in its own terms.

Again, we exempt $A$ from Einstein locality because we have deliberately left out its causal antecedents from our model.  With three-valued truth valuations, the argument works exactly as before: signalling would mean that $B$ is weakly correlated to $A$, whereas EL implies that there is an event $C$ associated to the past of $\cA$ such that $\tgamma(B)=\tgamma(C)$ $\forall \, \tgamma \in \tTheta$.  But in that case $C$ must also be weakly correlated to $A$, which violates our freedom of settings assumption.  So superluminal signalling is, unsurprisingly, banned in any three-valued theory obeying EL in the above sense.  For the following argument, it is crucial to note that this argument for a ban on superluminal signalling refers only to the three-valued version of EL and the three-valued meaning of correlation and weak correlation; we did not have to mention the ontic level at all.

So far, this discussion should have been quite comfortable even for those who are most skeptical of altering logic.  After all, this discussion has been nothing more than a rewrite of the standard situation to formalise the case in which some truth values are unknown.  Of course, this rewrite alone cannot save us from Bell's theorem\footnote{To prove Bell's theorem in this context we would still need to give some idea of how to put probabilities on our histories.  Assuming ontic definiteness, it suffices to say that we would put probabilities only on the ontic histories, or alternatively that if we did entertain probabilities for events to be intermediate, these, and all their logical combinations, would be entirely independent of the probabilities on the ontic histories.  Bell's theorem can then be derived as usual.}.  Thus, adding a third truth value does not, on its own, go beyond the standard picture.  The real break with the standard picture is yet to come. In order to maintain a satisfying causal principle even when faced with Bell's theorem, we must reject the principle of ontic definiteness.

\subsection{The break with the standard picture}

What does it mean to reject ontic definiteness? It means that our theory is an three-valued theory $\{\tOm,\alg,\tTheta,\cM,\Delta \}$ as before, but that the truth valuations in $\tTheta$ need no longer contain the two-valued truth valuations corresponding to the histories in $\Theta$ from which they were derived.  In other words, we no longer maintain that, although we do not know all truth values, nonetheless ``in reality'' all events have a definite truth value that we are simply ignorant of.  We accept that, in some circumstances, there may be no meaning to hypothesising about the truth or falsity of certain propositions.  To remind us of this change in view, we will rename the intermediate truth value ``indefinite''.  However, in order to avoid logical chaos, we have retained the other rules given above for the three-valued truth valuations.  In particular, in order to be an acceptable three-valued truth valuation, it must be possible to derive the truth valuation from members of some standard history space $\Om$ in the way described in section \ref{s:thirdvalue}. The point can be illustrated with a simple example.

Let us start by assuming ontic definiteness.  In this simple theory there are four allowed ontic histories.  Event $A$ corresponds to the proposition ``the box is open'', while event $B$ corresponds to the proposition ``the ball is in the box''.  The four possible histories cover every combination of truth and falsity here.  Thus, two of the histories, $A^c \cap B$ and $A^c \cap B^c$, correspond to ``the box is closed and the ball is inside'', and ``the box is closed and the ball is not inside''.  Finding the derived epistemic theory to go along with this ontic theory gives us histories signifying things like ``the box is closed and it is intermediate whether or not the ball is inside'' and so on.

Now let us reject ontic definiteness.  We are now free to imagine a theory in which it is in principle impossible to know whether the ball is in the box when it is closed.  If we want such a theory, we can consider just three histories: ``the box is closed and it is indefinite whether or not the ball is inside'', ``the box is open and the ball is inside'', and ``the box is open and the ball is not inside''.  These can be called the ``underivable truth valuations'', or more simply the ``basic truth valuations'', which now play a role similar to that which the ontic histories played before\footnote{There seem to be two ways of thinking here.  We can think of the basic truth valuations ontologically, in which case we might be tempted to say that the intermediate events in the basic truth valuations are ``indefinite in reality'', \textit{i.e.}~that the truth value $\hf$ as it appears in the basic truth valuations is \textit{ontologically distinct} from 0 and 1.  On the other hand, we can think of them epistemically, in which case we reject the idea that there are is anything like an ontic history at all, and give up the idea of the basic truth valuations as possible ``mirrors of reality'' (something similar to Dummet's ``antirealism'' position), in which case we may prefer to say ``unknown'' rather than ``indefinite''.  The distinction can be made -- or considered meaningless -- according to the taste of the reader as it has no effect on any of the following arguments.  The important thing is that the ``card game'' described previously works just as well with or without ontic definiteness.}: all other histories in the theory are derived from them in the way described in section \ref{s:thirdvalue}.  But now, if the box is closed, no allowed truth valuation gives a definite truth value to the event $B$.

This might bring Shr\"odinger's cat to mind.  However, rejecting ontic definiteness does not imply that we \textit{must} treat \textit{all} boxes in the way described here, but merely that we are liberty to describe \textit{some} cases like this.

The most important thing to note here is that \textit{our accounts of Einstein Locality and the ban on superluminal signalling still make perfect sense even if we drop ontic definiteness}.  Thus we see that objections (1) and (2) against altering logic given in section \ref{s:objections} are unfounded in this case.  By rejecting ontic definiteness we do no damage at all to the foundations on which the principle of causality is built.  Instead we have, so to speak, pulled away the table cloth while leaving the dinner set untouched.

Is there any reason to preserve ontic definiteness?  To argue that it cannot be sacrificed, one would have to come up with a reason why it is a useful or necessary principle.  However, ideas very similar to removing ontic definiteness (such as rejecting the law of the excluded middle in formal logic) have often been considered by philosophers and mathematicians \cite{Dummett:1963a,Brandom:1980a}\footnote{In their discussion of ``schematic worlds'', Rescher and Brandom come particularly close to the scheme considered here, especially if the truth-value $/P/$ defined in section 8 of \cite{Brandom:1980a} is compared to $\tgamma(P)$ as used above.}.  Besides this, it is commonly held that quantum mechanics is telling us that not all the properties of a system can be known even in principle; dropping ontic definiteness while preserving the logical structures presented above is a particular formalisation of this view.

This new picture is certainly different from more na\"ive versions of realism.  But, it can be argued, it has all the attributes that those who call themselves realists (at least when it comes to quantum mechanics) would desire.  Firstly indefinite logic does not imply operationalism.  That is, it does not rule out the attribution of truth values to anything but operational events concerning preparations and measurements.  We can still infer definite truth values about non-operational events by applying such an indefinite theory.  So no special place need be given to the concepts of preparation or measurement.  To the contrary, applying Einstein Locality will often \textit{require} us to hypothesise events beyond the operational level to serve as explanations of correlations.  Secondly, in this new picture, truth is still universal: it is not the case here that a proposition that ``appears to be true'' to an observer may in fact hold some different truth value in the greater scheme of things.  There is only one universal truth value, $1$, signifying ``true''.  Reliable observers should agree on all definite truth values in the most obvious sense.  In these senses, indefinite logic provides all the objectivity that realists demand.

In addition, the new picture supplies something that previously seemed to be impossible:  even while maintaining a satisfying causal principle, consistency with the results of quantum mechanics is possible.

\subsection{The EPRB experiment}
\label{s:eprb}

We now turn to the EPRB experiment to see how rejecting ontic definiteness allows us to evade Bell's theorem.  In our model, there will be sixteen basic histories in $\tTheta$.  We have two spacelike regions $\cA$ and $\cB$.  Region $\cA$ contains a setting event $A_s$ and a result event $A_r$, and similarly for $\cB$, we have $B_s$ and $B_r$. For instance, in the standard EPRB scenario with spin-$\hf$ particles, $A_s$ could correspond to the proposition ``The Stern-Gerlach detector is set to angle $\theta$ rather than $\phi$'' and $A_r$ could mean ``the outcome is spin up''.  We also require that, within $\cA$, $A_s$ and $A_r$ are associated to two subregions $\cA_s\subset \cA$ and $\cA_r \subset \cA$, such that $\cA_s$ is to the past of $\cA_r$, and similarly for $\cB$.  This is done so that the setting counts as being to the past of the result.

In the mutual past $\cP=J^-(\cA) \cap J^-(\cB)$ we will include four more events, $P_{A0}$, $P_{A1}$, $P_{B0}$ and $P_{B1}$.  There are sixteen allowed truth valuations in $\tTheta$ corresponding to the 16 possible definite truth valuations on the settings and results in regions $\cA$ and $\cB$.  The truth values for the past events are decided as follows.  For truth valuations in which setting $A_s$ is valued true, \textit{i.e.}~when $\tgamma(A_s)=1$, $P_{A1}$ holds the same truth value as $A_r$ while $P_{A0}$ is indefinite; when $\tgamma(A_s)=0$, $P_{A0}$ holds the same truth value as $A_r$ while $P_{A1}$ is indefinite.  The situation for the $B$ variables is the same with $B$ substituted for $A$.  The truth values of these eight events are given for all of the sixteen histories in a table to be found in appendix \ref{a:appEPRB}.

Specifying these truth values does not specify the truth values for all logical combinations of these events, and so to give the whole truth valuation we need to supply more.  For every truth valuation in $\tTheta$, two of the events $P_{A0}$, $P_{A1}$, $P_{B0}$ and $P_{B1}$ have indefinite truth values (\textit{e.g.}~, when $\tgamma(A_s)=\tgamma(B_s)=1$, $\tgamma(P_{A0})=\tgamma(P_{B0})=\hf$) and so the truth values of some of their logical combinations are not given by specifying their own truth values.  Several specifications can be given here which are compatible with our physical conditions.  This issue is treated at length in appendix \ref{a:eprb}.

The following argument establishes that Einstein locality condition holds for this model.  Exempting setting events from consideration as usual, it remains to show that the outcome events $A_r$ and $B_r$ are determined by some past event.  It is not hard to verify that $\tgamma((A_s \cap P_{A1}) \cup (A_s^c \cap P_{A0}))=\tgamma(A_r)$ for all of the 16 basic truth valuations, and similarly for the $B$-labelled events.  Because the result events are definite, their logical combinations are determined by logical combinations of their causal antecedents as usual.  It is also apparent that freedom of settings holds.  No past event in the model is weakly correlated to $A_s$ or $B_s$ for all of the basic truth valuations.  That is, none of the past events holds the truth value 0 whenever $A_s$ holds the truth value 0, and similarly for $B$, as is easily seen in the table.  It is also true that no (non-trivial) logical combination of the past events is weakly correlated to any (non-trivial) logical combination of the setting events, as shown in appendix \ref{a:eprb}, where these issues are treated in more detail.

Putting a probability distribution on these indefinite truth valuations (whilst maintaining freedom of settings in the probabilistic sense\footnote{Here this should mean that the probability of any event in the past, given that it is definite, is independent of the probability of the settings.  A similar treatment of the GHZ experiment \cite{Greenburger:1989,Mermin:1990a} can be given with no reference to probabilities.  This will be left for future work.}) it is possible to maximally violate the CHSH inequalities.  This is clear from the fact that the model puts no restrictions at all on the probability distributions over the events in $\cA$ and $\cB$ apart from no-signalling.  Thus we can have maximal correlations -- correlations even stronger than those allowed by quantum mechanics can be achieved here -- while maintaining a causal principle that bans superluminal influence.

\section{Discussion}
\label{s:discussion}

\subsection{The ``meta-level'' objection}
\label{s:meta}

The meta-level objection referred to before, in this case, is that we have exploited the ``causal loophole'' or (more or less equivalently) allowed backward causation.  After all, when $\tgamma(P_{A1})=1/2$, the truth value of the setting event $\tgamma(A_s)$ can only be $0$.  Doesn't this mean that $P_{A1}$ is determining $A_s$?  This is true, \textit{if} one treats $\tgamma$ not as a truth valuation but as a list of variables describing an ontic history, and goes on to use standard logic to talk about these 16 histories.  More precisely, we can invent a new ``meta-algebra of events'' $\halg$ in which the ``meta-events'' correspond to the truth valuations of events given by the histories $\tgamma$.  For instance, in the above example ``$\tgamma(P_{A0})=1/2$'' is a meta-event.  We can introduce a standard, definite truth valuation on $\halg$, which we call a ``meta-history'' $\hgamma$, corresponding to each indefinite truth valuation $\tgamma$.  For instance, for all of the 16 meta-histories in our example, $\hgamma(\tgamma(P_{A0}=1/2))=1$ if and only if $\tgamma(P_{A0}=1/2)$, and so on.    This is what is meant by the ``meta-level'' -- in this way it is always possible to find a corresponding definite theory for any indefinite theory.  This meta-level objection can be leveled at any such change in logic\footnote{Or at least, any change in logic for which the truth of an event as determined by a reliable observer is universal.}.  On the meta-level, standard logic applies, and we can define theories just as before.  So, as before, one of the assumptions of Bell's theorem mentioned above must be violated.  How then could using such an altered logic bring anything new to the discussion of Bell's theorem?

Even looked at from this ``meta-level'' point of view, there is something to be said for the altered logic set out above.  There is still a set of conditions here that bans superluminal signalling -- the indefinite version of EL (which, it can be seen, is \textit{stronger} than the definite version at the meta-level), the indefinite version of freedom of settings (which is \textit{weaker} than the definite version at the meta-level) and the condition that the setting and outcome variables can have the values 0 or 1 but never 1/2.  Together, these ban superluminal signalling.  From this point of view what we have done is to strengthen Einstein locality in such a way that we can ban superluminal signalling.  So even if we take this meta-level stance, the worst that can be said is that we are utilising the freedom of settings loophole.  Still, a condition can indeed be stated which bans superluminal signalling and yet allows violations of Bell's inequalities, which is surely progress of the type we were looking for.  Apart from this, though, motivation for this new ``causal principle'' is somewhat opaque from the ``meta-level'' point of view.

While this is certainly a possible stance to take, I will argue that it is unreasonable to insist on this ``meta-level'' viewpoint.  There is no justification for the claim that the three-valued version of Einstein Locality is ``not really'' Einstein Locality, when it serves exactly the same function as the definite version and is derived on exactly the same grounds.  The same holds for the three-valued version of freedom of settings.  We have already seen that they are motivated in the same way as the standard, definite expressions, and hopefully they can be just as useful for the progress of physics.  Furthermore, the ``card game'' described above is intuitive and standard for three-valued theories \textit{with} ontic definiteness, and this game proceeds in exactly the same way when ontic definiteness is dropped; in the light of this there is no reason to retreat to a meta-level perspective in order to ensure that we can reason effectively.  Additionally, for any definite events (events for which we know the truth value) we can reason exactly as usual.  This provides a strong argument that the indefinite statements are just as worthy of their names as their definite cousins, and that this way of speaking can be just as meaningful, and just as useful -- in fact more so because we are no longer in contradiction with quantum mechanics\footnote{An argument for a similar claim is made at length in \cite{Brandom:1980a}.  There, it is argued that Rescher and Brandom's inconsistent semantics need not be considered as merely derivative from the standard one, but could just as easily be considered the basic case from which the standard one is derived, a stance the authors call ``the parity thesis'' (see in particular section 13 of \cite{Brandom:1980a}).}.

It is true that, when $\gamma(P_{A1})=1/2$, the setting event $A_s$ can only be false.  From the indefinite logic point of view there is nothing odd about that.  Technically it does not violate our freedom of settings requirement because $P_{A1}$ is not weakly correlated to $A_s$.  To gain an intuition about this, we can imagine a theory that contains 3 basic indefinite histories, corresponding to the statements ``the box is closed today and whether the ball was in it yesterday or not is indefinite'', ``the box is open today and the ball was in it yesterday'' and ``the box is open today and the ball was not in it yesterday''.  For this perspective there seems to be nothing strange in saying that an event that happened in the past (the ball being in the box yesterday) can be made definite or indefinite by an event that happens today (opening the box to check).  In particular, it would be unreasonable to claim that the later event was determined by the earlier one (or \textit{vice-versa}).  All that is happened is that, if the box is open, this gives us grounds to attribute a definite truth value to whether the ball was in there yesterday, which are otherwise absent.  Thus, it can be the case that one event (even a past event) is ``made definite'' by another without them being correlated.   This is simply not what we mean by correlation, in this context.

\subsection{The relation to counterfactual definiteness}

The connection to the idea of counterfactual definiteness is now apparent.  By dropping the assumption of ontic definiteness, we can revive this ``loophole''.  In the treatment of the EPRB experiment given above, it is indeed the case that no definite value is attributed to the results of the measurements that are not performed.  This is therefore an example of lack of counterfactual definiteness.  In this setting however, the lack of counterfactual definiteness is not vulnerable to Norsen's criticism:  he derives counterfactual definiteness from the assumption of local causality (note that he could just as easily have assumed the slightly stronger Einstein Locality to reach the same conclusion) and freedom of settings, but assuming ontic definiteness. Instead, we have dropped Norsen's (and Bell's) assumption of ontic definiteness, while preserving the indefinite versions of of Einstein locality and freedom of settings.  This provides a way to evade Norsen's argument that counterfactual definiteness is required by local causality.  It should, again, be stressed that making this move does not commit us to operationalism, against which Norsen directs his argument \cite{Norsen:2006a}.

\subsection{Relation to other programmes}

The idea that ontic definiteness can be dropped, allowing the preservation of a meaningful relativistic principle of causality without contradicting quantum mechanics, should be of interest generally in inquiries into the foundations of quantum mechanics.  It is also of specific interest to more than one programme for interpreting quantum mechanics.  On its own, dropping ontic definiteness does not provide an interpretation, but only a new way to evade Bell's theorem.  Nevertheless, as has been noted, one of the prime decisions to be made when interpreting quantum mechanics (whether in an operationalist way or not) is what assumption of Bell's theorem to drop.  Using the newly uncovered assumption to build a full interpretation is surely of interest.

The most obviously related programme is the $\Psi$-epistemic approach.  Here, the quantum state is viewed, not as reflective of an element of reality (\textit{i.e.}~not ontologically), but instead as expressing our state of knowledge about the system in question.  Spekkens has proposed a toy model relying on these ideas that qualitatively reproduces many interesting features of quantum mechanics \cite{Spekkens:2004a}.  He introduces a principle under which, in a particular well-defined sense, the truth values of only half of all events can ever be known at once, thus limiting the possible epistemic states to always be ``coarser grained'' than the ontic states.  This is moving very near to the direction this article has taken.  However, although the possible epistemic states are limited in the toy model, it is still assumed that ontic states exist, from which the epistemic states are derived as above in section \ref{s:thirdvalue}. As we have seen, jettisoning this assumption allows us to evade Bell's theorem.  It would be most interesting to see if Spekken's toy model can be improved by removing the assumption of ontic definiteness, allowing it to model \textit{all} interesting features of quantum mechanics, including the stronger-than-classical correlations.  For instance, perhaps the model could be altered to produce ``stabiliser-group quantum theory'' (a reduced version of standard quantum theory), which has been found to be very similar to the toy model \cite{Coecke:2010a}.  If this could be achieved, the new toy model would indeed be able to do better than the original one in modelling the strong correlations found in GHZ states \cite{Greenburger:1989,Mermin:1990a}.

Connections can also be found to the anhomomorphic logic program \cite{Sorkin:2006wq,Sorkin:2010kg,Dowker:2007kz}.  In this approach to quantum mechanics, as here, the algebra of events is Boolean as in standard logic, and their truth values are still in $\{0,1\}$, but the truth valuation need not be homomorphic (so, for example, if $A$ is false and $B$ is false it does not necessarily follow that $A \cup B$ is false).  The are many possible ways to weaken the rules of classical logic after abandoning homomorphic truth valuations.  Interestingly, the one that has turned out to be most useful for quantum mechanics, the ``multiplicative scheme'', bears strong similarities to the ideas set out above.  That is, there is a one-to-one map between truth valuations and subsets of the history space $\Om$, as above.  In the truth valuation corresponding to subset $X$, the event $A$ is true if and only if $A \supset X$.  This is also true for the derived three-valued truth valuations described above, as noted in section \ref{s:thirdvalue}.  The difference is that, in anhomomorphic logic there are only two truth values, and so all other events are designated false (whereas, in the derived three-valued truth valuations, only events $A$ such that $A^c \supset X$ are false, and the rest are intermediate).  We see here that for every truth valuation in anhomomorphic logic, it is easy to substitute an indefinite truth valuation.  The events given indefinite truth values are exactly those events $A$ for which, in the anhomomorphic ``co-event'', $A$ and $A^c$ are both deemed false\footnote{This line of thought also suggests that the above evasion of Bell's theorem could be re-written in the anhomomophic logic language.}.  Methods have already been proposed to derive the set of possible anhomomorphic truth valuations from quantum mechanical theories.  This supplies a ``ready-made'' interpretation of quantum mechanics with indefinite histories and no ontic definiteness.  However, one would still have to prove in each case that the theory obeyed Einstein locality (or the weaker relativistic causal principle hinted at near the end of section \ref{s:el}) in this interpretation.  As yet, the theory contains no general rule that explictly ensures this.

%This quantum interpretation is also tied to a particular interpretation of probability, which seems to be a costly step that is not necessitated by the lack of ontic definiteness, and could possibly be avoided while exploiting the new ``loophole''.

It is also natural to wonder about the connection to the topos-based approach to quantum interpretation pioneered by Butterfield, D\"oring and Isham \cite{Isham:1998jv,Doring:2007ib}.  This approach also alters logic, but similarities do not go much further.  In this approach, what an observer would normally call true should properly be considered only ``true to them'', or true only in their particular context.  This is in contrast to the case discussed above where there is one truth value meaning ``true'', $1$, which is universal.  This is the main reason why it is difficult simply to transpose the discussion of causal principles given here to the case of the topos-based approach.  The discussion of whether a satisfying causal principle can be maintained in that interpretation may, for this reason, have more similarities to the discussion for many-worlds approaches.

\section{Conclusion}

This paper began by noting that Bell's theorem seems to prevent us from maintaining a satisfying relativistic principle of causation.  When interpreting quantum mechanics, we seem to be faced with a choice between maintaining the principle of common cause, or relativistic causal structure.  Thus the road chosen here depends on whether or not we see causality as a necessary part of any satisfying theory.

In this article, it has been shown that there is a hidden assumption that forces us to make this choice.  Ontic definiteness is the principle that, whether we can (even in principle) have grounds to know them or not, every proposition about physical systems must have a definite truth value, ``true'' or ``false''.  We have seen that this principle in no way underpins either the causal principle of Einstein Locality, or the assumption of freedom of settings used in Bell's theorem.  However, it does underpin Bell's derivation of a contradiction with the predictions of quantum mechanics, starting with these principles.  We are left with a choice: we can either abandon the demand for a satisfying relativistic causal principle, or we can abandon the principle of ontic definiteness.  Abandoning ontic definiteness seems to do no damage to the demands made by physicists who defend realism: we are still allowed, and moreover required, to make inferences about facts beyond the operational level.  On the other hand, those who reject approaches like deBroglie-Bohm pilot wave theory on the grounds that they contradict locality should find nothing to object to either.

There is a strong case to be made that, if we have to do without one, maintaining ontic definiteness is less desirable than maintaining a satisfying causal principle.  Arguments have been given above that no other way of evading Bell's theorem allows a satisfying causal principle to be maintained.  When interpreting quantum mechanics, therefore, dropping ontic definiteness is not only a new option, but is arguably the most attractive one available.

This raises the question of what such an interpretation would look like.  Above, it was suggested that simple models in the spirit of Spekken's toy model could be constructed to gain a better understanding of this.  One of the major questions here concerns the strength of correlations in quantum theory.  Without ontic definiteness, Einstein locality allows the strongest non-signalling correlations that are logically possible, while quantum mechanics places restrictions on the strength of correlations.  It is very interesting to ask whether this can be explained as the result of any simple set of physical principles.  These issues are of current interest for those studying foundational issues in quantum information, for example.  The same question arises for this new programme.  If Einstein locality alone does not imply the quantum restrictions on correlations, will it do the job when supplemented by other physical principles, and if so, what are they?  Hopefully, toy models can shed more light on this issue.

Other interesting extensions include the application to quantum gravity.  It is arguable that a proper understanding of quantum mechanics is a prerequisite for this.  But even so, it is interesting to ask at this stage what the principle of Einstein Locality would look like if we removed the assumption of a fixed background spacetime.  With general covariance in mind, this may require answers to some deep questions, along the lines of ``what is a region?''.  Perhaps addressing the issue in the conceptually simpler context of causal sets will lead to progress here.

There are many other avenues for future work.  One is to study Mermin's formulation of the GHZ experiment \cite{Mermin:1990a} in the light of the ontic definiteness assumption.  Beyond this, it would be of use to study indefinite logic independently of the derivation from standard logic, and to consider the role of probabilities more carefully.

We have seen that principles of causation, and the physicist's notion of realism, are closely intertwined concepts.  By further studying the consequences of rejecting ontic definiteness, it should be possible to make further progress in the task of restoring a satisfying picture of the microworld to our physical theories.

\section*{Acknowledgements}
The author is grateful to Jeremy Butterfield, Adam Caulton, Fay Dowker, Adrian Kent and Rafael Sorkin for useful discussions, and to Rob Spekkens for pointing out the work of Travis Norsen on counterfactual definiteness.

\begin{appendix}

\section{Causal Principles: what are they good for?}
\label{a:causal}

\subsection{Defending the causal principle}

In section \ref{s:bell} it was asserted that finding a satisfying causal principle that is compatible with quantum theory may help us to make scientific progress.  A few comments by way of clarification and justification of this statement are given in this appendix.

Causation furnishes us with an important mode of explanation.  Requiring causal explanation often means requiring the hypothesis of new events (in the sense in which the word is used above) to serve as common causes for otherwise unexplained correlations.  This is part of the story of discovering new substances and processes.  As a simple example, consider a case in which the movement of one tree is correlated to the movement of another tree.  Let us assume that we have good reason to think that the movement of one tree is not in itself causing the movement of the other.  In this case, it would be strange to explain the simultaneous movement of two trees by stating that it is in the nature of trees to move simultaneously, and leave it at that.  Instead, the PCC requires us to posit a new effect that furnishes the common causes for the simultaneous movement of trees.  In this way we enrich the subjects of our physical inquiry from the initially apparent events (which all have to do with the properties of trees) to new events (having to do with the wind).

Indeed, having gained confidence in our wind theory, we make no fundamental division between the events we originally categorised as ``apparent'' and those which we inferred from them using the causal principle.  The boundaries of what is considered operationally ``apparent'' and what is not can also change if we take the causal principle seriously: for example the electromagnetic field, at first taken as inferred through its effects on matter, is in modern theories no more or less deserving of a special ``apparent'' status than matter fields.  It is interesting to note that taking a strictly operational point of view (though it might be useful as a means of shedding unnecessary assumptions present in previous theories) prevents this kind of advance, by fixing the divide between the events that are considered apparent and the rest.  Thus, the causal principle is closely intertwined with the realistic (\textit{i.e.}~non-operational) world-view.  Moreover, it is interesting to ask: is there any use in insisting on a non-operational point of view \textit{without} also insisting on a satisfying causal principle?  In quantum mechanics we may well be able to solve the measurement problem by denying the PCC.  But in so doing, we give up the possibility of enriching the subjects of our physical inquiry by using a causal principle.

\subsection{Banning superluminal signalling is not enough}
\label{a:ban}

From an operational standpoint, it is tempting to ask: if we want a causal principle that bans superluminal signalling, why not simply take the ban on superluminal signalling itself to be the causal principle?  However, there is a good reason to make a distinction between the two ideas.  In the example of the previous section, we might shake one of the trees, and observe that distant trees are not caused to move.  This certainly adds to the evidence that there is no direct causal link between the movement of the two trees.  But it does not do anything to generate an \textit{explanation} of why the trees' movement is correlated despite this, leading to our discovery of the effect of wind.  A true causal principle, like the PCC, can do this for us.  As argued in the previous section, one of the points of reinstating a causal principle is to be able to talk about events that are not directly observable (or not at first considered to be), and so it defeats the object to retreat to an operational point of view .

There is another related reason not to accept the ban on superluminal signalling as a satisfying causal principle: doing so would confuse superluminal \textit{influence} with superluminal \textit{signalling}.  For example, imagine that we often observed a bright light being emitted from every point on a spacelike path through spacetime.  When trying to design a physical theory that explained this effect, it would be very unnatural to aim for a theory in which there was no superluminal influence.  But it could still be the case that these effects could not be controlled and used to signal.  A reasonable theory here could maintain a ban on superluminal signalling, but not a ban on superluminal influence in general.  It is superluminal influence in the more general sense that is banned by a satisfying relativistic causal principle.

\subsection{The challenge of causal antifundamentalism}

Some recent arguments have been made that the idea of a causal principle should, at this stage in the development of science, be rejected as a universal principle for science.  This is not the converse of the ``modest assertion'' made at the beginning of section \ref{s:bell}, and so need not concern us too much, but the claims still deserve some discussion in this context.  Butterfield has also discussed Norton's stance in the context of relativistic causality and quantum mechanics in \cite{Butterfield:2007b}.

Arthur Fine questions whether correlations need explanation \cite{Fine:1989a}.  Essentially, he is advocating dropping the PCC as a response to the problems of quantum mechanics.  Hopefully the arguments of the previous subsection are sufficient to give question to this.  John Norton gives a more historically-based criticism of causation as a fundamental principle of science \cite{Norton:2003a}.  He argues:

\begin{quote}
Mature sciences, I maintain, are adequate to account for their realms without need of supplement by causal notions and principles. The latter belong to earlier efforts to understand our natural world, or to simplified reformulations of our mature theories, intended to trade precision for intelligibility.  In this sense I will characterize causal notions as belonging to a kind of folk science, a crude and poorly
grounded imitation of more developed sciences.
\end{quote}

One part of Norton's argument is that our idea of causation has had to be redefined after major developments in physics, leading one to suspect that there exists no general form for what he calls ``causal notions'' at all.  He also argues that some examples even in Newtonian mechanics show violations of causal principles.  Finally he brings up the very real problem of Bell's theorem.  If Norton's position, which he dubs ``causal antifundamentalism'', is correct, we can conclude that theories should not be considered inadequate because they fail to obey a causal principle (like a version of the PCC), since causal principles are simply folk-science and may have already become obsolete.

Norton's position deserves serious consideration.  In particular it is hard to argue that the casual principle must be defined once and for all, and preserved through every possible development in science.  Nevertheless, I will argue that it is worthwhile to preserve a well-defined causal principle for the time being.

To the first point, the fact that ``causal notions'' have had to be refined, and that, at some points in history, our best understanding of them has been considered problematic, need not suggest their eventual dissolution.  This has not, for example, been the case for ``gravitational notions''.  Neither are the pathological examples given from Newtonian mechanics very worrying for those who seek to reinstate the idea of a causal principle in modern physics.  To start with the examples are based on highly unphysical examples (in one example, a perfect ball must be positioned at an exact point on an object with an infinitely precise profile).  They can, no doubt, easily be removed by appending reasonable and near-trivial restrictions to Newtonian mechanics, rendering the theory (non-relativistically, deterministically) causal. More importantly for practical purposes (although this does not contradict Norton), the use of causal arguments may yet have been important to the development of various theories despite their failure in pathological cases.  The remaining argument is that we have managed to make huge advances in quantum mechanics, a theory in which the causal principle is still mysterious.  This is certainly an important point to consider.  But it proves neither that there is no satisfying causal principle to be found here, nor that it is not advisable to find and use it.

Norton's argument that modern theories do not \textit{need} to be supplemented by causal principles in order to function does not affect my argument that causal principles can be a useful tool for understanding them, in order (amongst other things) to develop yet more powerful theories.  One explanation of correlations between events is that our theory allows (or requires) such correlations.  But this presupposes that such a theory exists.  This ``static'' view of theory ignores a crucial role of causation as a means of developing theories, as the tree example elucidates.

Finally, it is interesting to wonder why, if it is possible, anyone would object to reinstating a satisfying causal principle in quantum mechanics.  Before Bell's theorem was known, this was a possibility that had yet to be checked.  Objecting even to attempts to reinstate the causal principle in these circumstances seems quite perverse.  Bell himself used the analogy of ``the glove left at home'' to explain this:
\begin{quote}
If I find that I have bought one glove, and that it is right-handed, then I predict confidently that the one still at home will be seen to be left-handed.  But suppose we had been told, on good authority, that gloves are neither right- or left-handed when not looked at.  Then that, by looking at one, we could predetermine the result of looking the other, in some remote place, would be remarkable.  Finding that this is so in practice, we would very soon invent the idea that gloves are already one thing or the other even when not looked at.  And we would begin to doubt the authorities that had assured us otherwise.
\end{quote}
\cite{Bell:1990} Indeed, this is the substance of the EPR debate, as Bell notes.  Elsewhere he is quoted as saying of the EPRB experiment:
\begin{quote}
For me, it is so reasonable to assume that the photons in those experiments carry with them programs, which have been correlated in advance, telling them how to behave.  This is so rational that I think that when Einstein saw that, and the others refused to see it, \textit{he} [not Bohr] was the rational man... so for me, it is a pity that Einstein's idea doesn't work. The reasonable thing just doesn't work.
\end{quote}
\cite{Bernstein:1991} Here, the view set out in the present paper is that Bell is, although not in so many words, mourning the loss of a satisfying relativistic causal principle to modern physics due to its apparent incompatibility with the predictions of quantum mechanics.  Above, it is shown that Bell's impasse can be avoided by dropping the assumption of ontic definiteness.  It is true that dropping this principle is a greater leap than Einstein initially foresaw when he proposed that quantum mechanics should be causally ``completed'' while respecting relativistic ``locality''.  But on the other hand, since a road towards Einstein's goal is now open, it seems unreasonable to advocate against pursuing it.

\section{Comparing causal principles}
\label{a:comparing}

\subsection{The two formulations of Einstein locality}
\label{a:twodefs}

In section \ref{s:el} it was claimed that the two definitions of Einstein Locality that were given were equivalent.  This claim is proved in this appendix.

First let us assume the first formulation and derive the second.  For every history $\gamma_\cP^* \in \Omega|_\cP$ of events in the past region $\cP$, there exists a history $\gamma_\cR^* \in \Omega|_\cR$ for $\cR$ such that for all $\gamma \in \Theta$, if $\gamma|_\cP=\gamma_\cP^*$ then $\gamma|_\cR=\gamma_\cR^*$.  This implies the following:

\begin{equation}
\label{e:det}
\forall \gamma \in \Theta, \gamma' \in \Theta, A \in \alg|_\cR, \gamma|_\cP=\gamma'|_\cP \Rightarrow \gamma(A)=\gamma'(A) .
\end{equation}
% for all $\gamma_\cP^* \in \Omega|_\cP$ and for all events $A \in \alg|_\cR$, there exists $t\in\{0,1\}$ such that if $\gamma|_\cP=\gamma_\cP^*$ then $\gamma(A)=t$ for all $\gamma \in \Theta$.

In words, all allowed histories that match on $\cP$ give the same truth value to all events associated to $\cR$.  This allows us to construct an event associated to $\cP$ that shares the same truth value as $A$ for all allowed histories:

\begin{equation}
\label{e:b}
B=\{\gamma \in \Omega \, : \gamma|_\cP=\gamma'|_\cP \Rightarrow \gamma'(A)=1 \, \forall \gamma' \in \Theta\}.
\end{equation}

For any $\gamma_\cP^* \in \Omega|_\cP$, the event $\{\gamma \in \Omega \, : \gamma|_\cP = \gamma_\cP^*\}$ is associated to $\cP$ by the definition of $\Omega|_\cP$, and the event $B$ is associated to $\cP$ by virtue of being the union of a set of such events.  It remains to show that the event $B$ is correlated to $A$.  Consider an allowed history $\gamma \in \Theta$ such that $\gamma(A)=1$.  Any such history will fulfil the condition in (\ref{e:b}) due to (\ref{e:det}), and so it will be in $B$, giving $\gamma(B)=1$ in this case.  Any $\gamma \in \Theta$ such that $\gamma(A)=0$ will fail the condition of (\ref{e:b}) for the same reason, and will not be in $B$, giving $\gamma(B)=0$.  Thus $\gamma(A)=\gamma(B)$ for all $\gamma \in \Theta$, or in other words, they are correlated.  For an arbitrary event $A \in \alg|_\cR$, we have found an event $B \in \alg|_\cP$ that is correlated to $A$, and thus satisfied the second definition of Einstein locality.

Now let us consider the converse case.  We start with the assumption that $\forall A \in \alg|_\cR$, $\exists B \in \alg|_\cP$ such that $\gamma(A)=\gamma(B)$ $\forall \gamma \in \Theta$.  Consider an event $A=\{\gamma \in \Omega \, : \gamma|_\cR = \gamma_\cR^*\}$ for some $\gamma_\cR^* \in \Omega|_\cR$.  This event is associated to $\cR$.  Note that for all histories $\gamma \in \Omega$, $\gamma|_\cR=\gamma_\cR^*$ if and only if $\gamma(A)=1$.  By the second formulation of EL, there is an event $B \in \alg|_\cP$ correlated to $A$.  In other words, for all histories $\gamma \in \Theta$ if $\gamma(B)=1$ then $\gamma(A)=1$, and so $\gamma|_\cR=\gamma_\cR^*$.  The history restricted to $\cP$, $\gamma|_\cP$ fixes the truth value of $B$.  Also, for every history $\gamma \in \Omega$ there exists some $\gamma_\cR^*$ such that $B$ is true.  Thus, for every restricted history $\gamma_\cP^* \in \Omega|_\cP$, there exists a history $\gamma_\cR^* \in \Omega|_\cR$ for $\cR$ such that for all $\gamma \in \Theta$, if $\gamma|_\cP=\gamma_\cP^*$ then $\gamma|_\cR=\gamma_\cR^*$.  We have recovered the first formulation of Einstein locality.

\subsection{The two formulations of the Nonprobabilistic PCC}
\label{a:twoPCCs}

In \ref{s:el} two formulations of a non-probabilistic principle of common cause are given, called NPPCm and NPCCj.  These formulations are claimed to be equivalent.  These claims are proved in this section.

The context is a nonprobabilistic theory defined by $\{\Omega,\alg,\Theta,\cM,\Delta \}$, two spacelike regions $\cA$ and $\cB$, their ``joint past'' $\cP_j=J^-(\cA) \cup J^-(\cB) \backslash (\cA \cup \cB)$ as shown in figure \ref{f:pcc} and their ``mutual past'' $\cP_m=J^-(\cA) \cup J^-(\cB)$.  The theory obeys the NPCCj if the following principle holds: for any pair of events $A \in \alg|_\cA$ and $B \in \alg|_\cB$, if $A$ and $B$ are correlated, then there exists a ``common cause'' event $C \in \alg|_{\cP_j}$ that is correlated to both $A$ and $B$.  The definition of NPCCm is the same with $\cP_j$ replaced by $\cP_m$.

First, it is easy to see that NPPCm implies and NPCCj.  NPPCm requires that all pairs of correlated events $A$ and $B$ as described above have common cause event in the mutual past $\cP_j=J^-(\cA) \cap J^-(\cB)$.  This is a subset of the joint past $\cP_m=J^-(\cA) \cup J^-(\cB) \backslash (\cA \cup \cB)$.  Thus, if there is a common cause event in the mutual past then there is also one in the mutual past, and NPCCm is also satisfied.

Conversely, consider the regions $\cY=J^-(\cA) \backslash J^-(\cB)$ and $\cZ=J^-(\cB) \backslash J^-(\cA)$.  Because $\cA \subset \cY$ and $\cB \subset \cZ$, $\alg|_\cA$ is a subalgebra of $\alg|_\cY$ and $\alg|_\cB$ is a subalgebra of $\alg|_\cZ$.  Thus $A \in \alg|_\cY$ and $B \in \alg|_\cZ$.  If the joint past version of the NPCC holds for all pairs of spacelike regions, then it must apply to $\cY$ and $\cZ$.  The joint past of $\cY$ and $\cZ$ is $\cP_m=J^-(\cA) \cap J^-(\cB)$.  Applying the joint past version of the NPCC we see that, if $A$ and $B$ are correlated, then there exists an event $C \in \alg|_{\cP_m}$ that is correlated to both $B$ and $C$.  Thus the mutual past version of the NPCC also holds.  This proves that NPCCm follows from NPCCj, and as a corrolary that NPCCm and NPCCj are equivalent.  This result, and the proofs, are similar to those already known for the probabilistic case \cite{Henson:2005wb}.

This result may seem questionable on first sight.  Why can't the entire chain of causal antecedents of $A$ lie in $J^-(\cA) \backslash \cA$ but outside of $J^-(\cB)$, and similarly for $B$, so that there is no common cause in the mutual past? Yet we have proved that the NPCCm principles follows from NPCCj, which is justified on general grounds.  Could there be something wrong with our assumptions, therefore?  But we should consider what it means for the entire chain of causal antecedents of $A$ to lie outside of $J^-(\cB)$.  If there is an initial cosmological spatial hypersurface beyond which no past exists, this would imply that there are two events on that surface (one the causal antecedent of $A$, and the other of $B$), at spacelike positions, that are correlated to each other.  NPCCj disallows these ``initial correlations'' as they have no common cause (similarly ``correlations that propagate in from past infinity'' are banned).  This point is discussed in \cite{Henson:2005wb} for the probabilistic case.

\subsection{Einstein locality and screening off}
\label{a:ELandSO2}

It was also claimed in section \ref{s:el} that Einstein locality implied a certain stochastic relativistic causality condition when a probability distribution was put on $\Theta$.  This is now shown.  Along similar lines, Bell comments that a deterministic version of causality follows from his local causality when probabilities are restricted to be 1 and 0 \cite{Bell:1990}.

Consider two spacelike spacetime regions $\cA$ and $\cB$, their union $\cC=\cA \cup \cB$, their joint past region $\cR=J^-(\cA \cup \cB)$ and their exclusive past $\cP=\cR \backslash (\cA \cup \cB$), as shown in figure \ref{f:pcc}.  According to the first formulation of Einstein locality, for every history $\gamma_\cP^* \in \Omega|_\cP$ of events in the past region $\cP$, there exists a history $\gamma_\cR^* \in \Omega|_\cR$ for $\cR$ such that for all $\gamma \in \Theta$, if $\gamma|_\cP=\gamma_\cP^*$ then $\gamma|_\cR=\gamma_\cR^*$.  Since $\cC$ is a subregion of $\cR$, it also holds that for every history $\gamma_\cP^* \in \Omega|_\cP$ of events in the past region $\cP$, there exists a history $\gamma_\cC^* \in \Omega|_\cC$ such that for all $\gamma \in \Theta$, if $\gamma|_\cP=\gamma_\cP^*$ then $\gamma|_\cC=\gamma_\cC^*$.  Similar statements also hold for the regions $\cA$ and $\cB$.

Consider a probability distribution $\mu$ on $\Omega$ with support in $\Theta$, and assume Einstein Locality.  Consider also the event $P=\{\gamma \in \Omega \, : \gamma|_\cP = \gamma_\cP^*\}$, which we can call a ``full specification'' of $\cP$.  Conditioning on $P$ is equivalent to fixing the history restricted to $\cP$ to be $\gamma_\cP^*$.  According to the first formulation of EL, this also fixes the history on $\cC$ to some value $\gamma_\cC^*$, and hence fixes the the history in $\cA$ and $\cB$, making all probabilities conditioned on $P$ either 0 or 1.  This means that, for all events $C$ associated with $\cC$, $\mu(C|P)=0$ if $\gamma_\cC^*(C)=0$ and $\mu(C|P)=1$ if $\gamma_\cC^*(C)=1$.  For any pair of events $A$ associated to $\cA$ and $B$ associated to $\cB$, the fact that $\gamma_\cC^*$ is a homomorphism then implies that $\mu(A|P)\mu(B|P)=\mu(A \cap B |P)$.  In other words, we have shown that all pairs of events $A$ and $B$ (associated to regions $\cA$ and $\cB$ respectively) are statistically uncorrelated when conditioned on any full specification of events in their joint past region $\cP$.  This ``screening off'' condition is called SO2 in \cite{Henson:2005wb}.  This SO2 condition has been shown to be equivalent to the similar condition (called SO1) in which the ``mutual past'' $J^-(\cA \cap \cB)$ is substituted for the joint past in the definition \cite{Henson:2005wb}, which is essentially the same as one of Bell's formulations of local causality (given in ``The theory of Local Beables'' \cite{Bell:1987}).

\section{Comparison with Reichenbach's logic}
\label{a:reichenbach}

In several important senses, the logic introduced in section \ref{s:thirdvalue} has a precursor in Reichenbach's three-valued logic \cite{Reichenbach:1944a}.  Reichenbach also identified problems with standard treatments of quantum mechanics, amongst them ``causal anomalies'', and sought to remedy them  by introducing a third truth value which he termed ``indeterminacy''.  However, there are important differences between the approaches.  Firstly, Reichenbach's work on three-valued logic was prior to Bell's on non-locality, and to Reichenbach's own later work on the principle of common cause.  His discussion has nothing explicit to say about the problem of restoring the PCC in the face of Bell's theorem, therefore.  Moreover, the scheme given above is not a simple extension of Reichenbach's ideas to solve the problem in hand; there are also significant technical differences.

In the approach above, the normal Boolean algebra of events in retained, and only the normal logical operations are extended to the three-valued logic.  In contrast, Reichenbach allows completely general logical operations, mapping the ``input'' truth values of events and pairs of events to the ``outputs'' in all possible combinations.  For example, the logic employed above implies the rule $\tgamma(A)=1 \Rightarrow \tgamma(A^c)=0$, $\tgamma(A)=\hf \Rightarrow \tgamma(A^c)=\hf$, $\tgamma(A)=0 \Rightarrow \tgamma(A^c)=1$.  For Reichenbach this is only one of the possible logical operations with one argument.  As well as this ``diametrical negation'' which he terms $-A$ he considers ``cyclical negation'' $\sim A$ that gives $\tgamma(A)=1 \Rightarrow \tgamma(\sim A)=\hf$, $\tgamma(A)=\hf \Rightarrow \tgamma(\sim A)=0$, $\tgamma(A)=0 \Rightarrow \tgamma(\sim  A)=1$ (our set theoretic notation, which implies a Boolean algebra of events, fails here, but hopefully the meaning remains clear).  Likewise there are novel two-argument logical operations in Reichenbach's approach.  Every logical operation of this type is allowed.  Thus, there are events in the three-valued logic corresponding to all the events at the meta-level described in section \ref{s:meta} (\textit{i.e.} that have the same definite truth value as the given meta-level event).  Given this, Reichenbach's approach does not allow the strategy used above to give a causal explanation of the EPRB experiment: using the novel logical operations, an event can be defined in the past that is correlated with the setting events, violating our assumptions.  Consider the model of the EPRB experiment given in section \ref{s:eprb}. Defining a logical operation ``is definite", $?A$, such that $\tgamma(A)=\hf \Rightarrow \tgamma(?A)=0$, $\tgamma(A)=0$ or $1 \Rightarrow \tgamma(?A)=0$, we find that the event $?P_{A1}$ determines the setting $A_s$.  Thus admitting Reichenbach's extra logical operations leads to a version of the ``meta-level'' objection discussed in section \ref{s:meta}.  The answers to the objection are the same.  Nothing forces us to allow these operations in our new logic; indeed, from the perspective given in the main text they appear to be an unnatural addition.

Another feature of Reichenbach's approach is that is it explicitly operationalist in character.  Thus he states ``it is possible to introduce an intermediate truth value which may be called \textit{indeterminacy}, and to coordinate this truth value to the group of statements which in the Bohr-Heisenberg interpretation are called \textit{meaningless}'' -- that is, propositions about unmeasured properties. With this operational meaning for the third truth value, the principle of common cause that Reichenbach later formulated cannot be restored, as argued in section \ref{s:droppcc}.

However, there are several points from Reichenbach's discussion that do carry over.  One is that complimentarity can be given a meaning based on truth values in the logical system.  Two events $A$ and $B$ may be considered complimentary if and only if, for all $\tgamma \in \tTheta$, $\tgamma(A)=\hf$ if and only if $\tgamma(B)\neq\hf$.  Thus events $P_{A0}$ and $P_{A1}$ in section \ref{s:eprb} are complimentary.  This observation is based on a point made by Reichenbach, although in this context the definition of complimentarity is expanded somewhat from its usual operational setting.

Reichenbach also recognises that introducing a third truth value may have an impact on the EPR ``paradox''.  After noting that Bohr's opinions on EPR ``do not seem to us to be stated sufficiently clearly to admit of an unambiguous interpretation'', Reichenbach confirms with EPR that ``what is proved by the existence of correlated systems is that it is not permissable to say that \textit{the value of an entity before the measurement is different from the value resulting in the measurement}'' (added italics) -- if, the context suggests, one assumes relativistic causal structure.  Reichenbach's proposed solution to the problem is that this italicised statement, though untrue, may not be false but instead hold the indeterminate truth value.  However, it is not clear that Reichenbach's proposal saves us from ``causal anomalies'', as he calls them, because it is not made clear exactly what constitutes a causal anomaly in the EPR setting, especially once the three-valued logic is introduced.  This clarification would have to wait for the later work of Reichenbach, and Bell, on causal principles.  The three-valued definition of Einstein Locality given in section \ref{s:ELrevisited} provides such a definition for a three-valued logic.  Using this definition, Reichenbach's proposed solution is not adequate, as he does not provide causal antecedents for the inarguably determinate events: the measurement outcomes\footnote{Perhaps the ``meta-level version'' of EL mentioned in section \ref{s:meta} would be more appropriate for Reichenbach's logic, but again this runs into trouble if the truth ``value of an entity'' in a distant region can be changed from $\hf$ to $1$ or $0$ by the free choice of an experiment.}.

\section{More details of the EPRB experiment}
\label{a:eprb}

This appendix gives some further explanation and discussion of points covered in section \ref{s:eprb}.

\subsection{More on free settings}

Before we come to discuss the details of the EPRB experiment, more needs to be said about the role of free settings.  In the EPRB experiment there is more than one setting, and so questions arise beyond those already addressed in sections \ref{s:signalling} and \ref{s:ELrevisited}.  For example, should logical combinations of settings also be treated as settings?  And should events that are logical combinations of settings and other events require causal antecedents?  We need to be more systematic about such questions to ensure that our model the EPRB experiment obeys freedom of settings and Einstein Locality for all events. We will start with the case of standard two-valued logic.

Setting events are a set of events in $\alg$.  We will say that logical combinations of settings are also settings, \textit{i.e.} the settings form a subalgebra of $\alg$ (this may seem odd in that the ``trivial events'' $\Omega$ and $\emptyset$ now qualify as settings, but this is the easiest way to proceed).

The main condition on settings is that they are not influenced by any past events in the model.  Consider a setting $A$ associated to region $\cA$.  If $A \cap \Theta$ is a proper subset of $\Theta$ (\textit{i.e.}~if there are histories in $\Theta$ for which $A$ is false), then there can be no event $C$ associated to $J^-(\cA) \backslash \cA$ such that $C \neq \emptyset$ and $C \cap \Theta \subset A \cap \Theta$.  In other words $C$ cannot be weakly correlated to $A$ except in trivial cases.  As well as this, settings associated to disjoint regions should not be weakly correlated to each other, a condition easily satisfied in the model of the EPRB experiment.

As well as these conditions on settings, the causal condition of Einstein Locality is weakened for settings.  This is because settings are supposed to be a ``free choice'', and thus whatever it is that determines them need not be relevant for the rest of the model (for example, if we decided the settings based on what we had for breakfast, it should not be necessary to include our breakfast in the model to get a good description of the system in question).  In order to exempt signals from Einstein Locality, we will weaken the second formulation of Einstein Locality given above.

As before we have a region $\cX$, its past $\cR=J^-(\cX)$ and its ``exclusive past'' $\cP=J^-(\cX) \backslash \cX$ as shown in figure \ref{f:el}.  The set $\alg|_\cP$ is the subalgebra of events associated to the past region.  Consider now the subalgebra $\alg_{\text{settings}}$ generated by $\alg|_\cP$ \textit{and} the setting events associated to $\cR$.

\paragraph{Einstein Locality (with settings):} Consider a nonprobabilistic theory defined by $\{\Omega,\alg,\Theta,\cM,\Delta \}$, and regions $\cX$, $\cR=J^-(\cX)$ and $\cP=J^-(\cX) \backslash \cX$. The theory is \textit{Einstein local} if the following condition holds for all regions $\cX$:

$\forall A \in \alg|_\cR$, $\exists B \in \alg_{\text{settings}}$ such that $\gamma(A)=\gamma(B)$ $\forall \gamma \in \Theta$.

\paragraph{} Previously, all causal antecedents had to be in the past region $\cP$.  Now we have expanded the allowed causal antecedents to also include setting events in the region $\cR$ itself.  Thus settings are exempted in the sense that they are allowed to be their own causal antecedents, which is a trivial requirement.  This also deals with events that are logical combinations of settings with other events.  Consider for example the event $A \cap B$ where $A$ is a free setting associated to region $\cA$ and $B$ is a non-setting event associated to a spacelike region $\cB$.  Here, $B$ itself requires a causal antecedent which we can call $D$, but $A \cap D$ is an acceptable causal antecedent for $A \cap B$, even though this event is not associated to $J^-(\cA \cup \cB) \backslash (\cA \cup \cB)$.

As in the main text, these rules are easily extended to the case of theories using the three truth values.  In this case, we still require the settings to form a subalgebra of $\alg$.  To prevent past influence on settings, we require the following (which follows for the derived three-valued theory from the condition above for two-valued theories, in the same way as in section \ref{s:ELrevisited}).  Consider a setting $A$ associated to region $\cA$.  If there exists  $\tgamma \in \tTheta$ such that $\tgamma(A)=0$, then there can be no event $C$ associated to $J^-(\cA) \backslash \cA$ such that, for all $\tgamma \in \tTheta$, $\tgamma(C)=0$ whenever $\tgamma(A)=0$, and that is nontrivial in the sense that there exists $\tgamma \in \tTheta$ such that $\tgamma(C)=1$. This is the three-valued version of the statement that $C$ cannot be weakly correlated to $A$ except in trivial cases.

Again, the statement of Einstein locality with settings is the same as the statement without settings, found in section \ref{s:ELrevisited} except with $\alg_{\text{settings}}$ replacing $\alg|_\cP$.  These conditions -- freedom of settings and Einstein locality with the exemption for settings -- are the ones that the EPRB experiment must obey.

\subsection{Application to the EPRB experiment}
\label{a:appEPRB}

In the model of the EPRB experiment provided in the text, 16 basic three-valued truth valuations are provided for the events of the model: the setting events $A_s$ and $B_s$, the results events $A_r$ and $B_r$, and the four $P$ variables assigned to the past.  The following is a table showing the truth values for these events in each of these 16 truth valuations:

\vskip 12pt
\begin{tabular}{| r || r | r | r | r | r | r | r | r |}
\hline
$i$ & $\tgamma_i(A_s)$ & $\tgamma_i(B_s)$ & $\tgamma_i(A_r)$ & $\tgamma_i(B_r)$ &
 $\tgamma_i(P_{A0})$ & $\tgamma_i(P_{A1})$ & $\tgamma_i(P_{B0})$ & $\tgamma_i(P_{B1})$ \\
\hline
%  i    As    Bs    Ar    Br   PA0   PA1   PB0   PB1
   1 &   0 &   0 &   0 &   0 &   0 & \hf &   0 & \hf \\
   2 &   0 &   0 &   0 &   1 &   0 & \hf &   1 & \hf \\
   3 &   0 &   0 &   1 &   0 &   1 & \hf &   0 & \hf \\
   4 &   0 &   0 &   1 &   1 &   1 & \hf &   1 & \hf \\
   5 &   0 &   1 &   0 &   0 &   0 & \hf & \hf &   0 \\
   6 &   0 &   1 &   0 &   1 &   0 & \hf & \hf &   1 \\
   7 &   0 &   1 &   1 &   0 &   1 & \hf & \hf &   0 \\
   8 &   0 &   1 &   1 &   1 &   1 & \hf & \hf &   1 \\
   9 &   1 &   0 &   0 &   0 & \hf &   0 &   0 & \hf \\
  10 &   1 &   0 &   0 &   1 & \hf &   0 &   1 & \hf \\
  11 &   1 &   0 &   1 &   0 & \hf &   1 &   0 & \hf \\
  12 &   1 &   0 &   1 &   1 & \hf &   1 &   1 & \hf \\
  13 &   1 &   1 &   0 &   0 & \hf &   0 & \hf &   0 \\
  14 &   1 &   1 &   0 &   1 & \hf &   0 & \hf &   1 \\
  15 &   1 &   1 &   1 &   0 & \hf &   1 & \hf &   0 \\
  16 &   1 &   1 &   1 &   1 & \hf &   1 & \hf &   1 \\
\hline
\end{tabular}
\vskip 12pt

This table, however, does not fully specify the sixteen truth valuations $\{\tgamma_i(\,\cdot)\}$.  What, for example, is the value of $\tgamma_1(P_{A1} \cup P_{B1})$?  The answer to questions like these does not follow from the values in the truth table above as it would for classical logic.  As explained in section \ref{s:thirdvalue}, to specify these it is sufficient to specify a set of definite truth valuations from which the history is derived.  Consider for example the following four definite histories:

\vskip 12pt
\begin{tabular}{| r || r | r | r | r | r | r | r | r |}
\hline
$i$ & $\gamma_i(A_s)$ & $\gamma_i(B_s)$ & $\gamma_i(A_r)$ & $\gamma_i(B_r)$ &
 $\gamma_i(P_{A0})$ & $\gamma_i(P_{A1})$ & $\gamma_i(P_{B0})$ & $\gamma_i(P_{B1})$ \\
\hline
%  i    As    Bs    Ar    Br   PA0   PA1   PB0   PB1
   1 &   0 &   0 &   0 &   0 &   0 &   0 &   0 &   0 \\
   2 &   0 &   0 &   0 &   0 &   0 &   0 &   0 &   1 \\
   3 &   0 &   0 &   0 &   0 &   0 &   1 &   0 &   0 \\
   4 &   0 &   0 &   0 &   0 &   0 &   1 &   0 &   1 \\
\hline
\end{tabular}
\vskip 12pt

Indefinite truth valuations derived from subsets of this set can produce the truth values given for $\tgamma_1(\, \cdot)$ above.  To give $\tgamma_1(P_{A1})=\tgamma_1(P_{B1})=\hf$ the subset must contain at least one definite truth valuation with each value for $P_{A1}$ and $P_{B1}$.  The sets $X=\{ \gamma_1(\, \cdot), \gamma_4(\, \cdot) \}$ and $Y=\{ \gamma_2(\, \cdot), \gamma_3(\, \cdot) \}$ have this property.  These correspond to saying that, while the truth values given to $P_{A1}$ and $P_{B1}$ are indefinite, there are still definite statements to be made about their logical combinations.  For example $\tgamma_Y(P_{B1} \cup P_{A1})=1$ as this event has truth value 1 for both $\gamma_2(\, \cdot)$ and $\gamma_3(\, \cdot)$.  The set $Z=\{ \gamma_1(\, \cdot), \gamma_2(\, \cdot), \gamma_3(\, \cdot), \gamma_4(\, \cdot) \}$ and all the three-member subsets also qualify, giving less definite events.  As an example $\tgamma_1(\, \cdot)$ can be set equal to $\tgamma_Z(\, \cdot)$, and similarly, each other allowed history $\tgamma_i(\, \cdot)$ can be derived from the set of four definite histories that share the definite truth values listed for $\tgamma_i(\, \cdot)$ in the first table above.

However, we can see that several sets of truth valuations following the table above will satisfy our conditions of Einstein locality and freedom of settings.  For the proof of freedom of settings, we will need to make the specification that for all sixteen of the indefinite truth valuations listed in the table, the conjunction of the two indefinite past events gives the same value $w$, which can be 0 or $\hf$.  In other words, $\tgamma_k(P_{A1} \cap P_{B1})=\tgamma_l(P_{A1} \cap P_{B0})=\tgamma_m(P_{A0} \cap P_{B1})=\tgamma_n(P_{A0} \cap P_{B0})=w$ for all values of the indices such that $1 \leq k \leq 4$, $5 \leq l \leq 8$, $9 \leq m \leq 12$ and $13 \leq n \leq 16$.  Similarly, unique values $x$, $y$, and $z$ must be given to the conjunctions of the indefinite past events and their compliments.  That is, $\tgamma_k(P_{A1} \cap P_{B1}^c)=\tgamma_l(P_{A1} \cap P_{B0}^c)=\tgamma_m(P_{A0} \cap P_{B1}^c)=\tgamma_n(P_{A0} \cap P_{B0}^c)=x$ under the same conditions on the indices as before, and similarly for $P_{A1}^c \cap P_{B1}$ and $P_{A1}^c \cap P_{B1}^c$ and $y$ and $z$.  Specifying these four variables specifies all of the truth values for each truth valuation.  While there are other specifications of the truth valuations consistent with the above table that would also obey our physical conditions, the main point of interest here is that even one exists.

The settings here are $A_r$ and $B_r$ but also all logical combinations of them such as $A_r \cap B_r$, $A_r \cap B_r^c$, and so on.  Firstly, it will be shown that Einstein locality in the ``settings version'' holds.  We do not require causal antecedents for the $P$ events (or any non-trivial logical combinations with them) as they are the earliest events that we have modelled\footnote{We can imagine a causal chain descending further into the past, but there is no need to model this.  If there were correlations between some of these ``initial events'' that were associated to spacelike regions, leaving out their causal antecedents would be more questionable, but we do not have to address this issue here since all past events are associated to the same region.}).  Thus we only need establish causal antecedents for events associated with the remaining region $\cA \cup \cB$. As mentioned in the main text, it can be seen from the table above that $A_r$ and $B_r$ have causal antecedents,  because $\tgamma((A_s \cap P_{A1}) \cup (A_s^c \cap P_{A0}))=\tgamma(A_r)$ for all of the 16 basic histories, and similarly for $B_r$.  The ``causal antecedents" of the settings are themselves.  All other events associated with $\cA \cup \cB$ are logical combinations of these events.  The causal antecedent of any logical combination of such events simply the same logical combination of the causal antecedents of these events.  This covers all events in the model, establishing Einstein Locality.

It remains to show freedom of settings.  For each of the setting events, the potential weakly correlated ``influencing'' events are the events that lie to the past: the four $P$ events and their logical combinations.  There must be no event $C$ associated to the past that has $\tgamma_i(C)=1$ for at least one value of $i$, and that also has $\tgamma_i(C)=0$ for all values of $i$ such that $\tgamma_i(A)=0$, for any setting event $A$ (apart from the trivial ``setting'' $\emptyset$).  We only need show that these conditions do not hold for any ``full specification'' of the settings, for example $A_s \cap B_s$, $A_s \cap B_s^c$ and so on.  If they do not hold for these events then they cannot be true of any disjunction of them, as a truth valuation cannot have $\tgamma_i(A_1 \cup A_2)=0$ if it is the case that $\tgamma_i(A_1)=0$ and $\tgamma_i(A_2)=0$.  Disjunctions of the full specifications cover all non-trivial logical combinations of the settings.  Let us consider the full specification $A_s \cap B_s$, which is true for the truth valuations for which $13 \leq i \leq 16$ in the table and false for $1 \leq i \leq 12$.

Now let us consider full specifications of the past events, meaning the sixteen ``finest-grained'' events such that knowing that one of these events had truth value 1 would imply the truth value of all other past events.  These are the events $P_{A0} \cap P_{A1} \cap P_{B0} \cap P_{B1}$, $P_{A0} \cap P_{A1} \cap P_{B0} \cap P_{B1}^c$ and so on.  Let us label these events $\{ F_j \}$ where $j$ runs between one and sixteen ($j$ is used to distinguish this index from the index $i$ labelling the sixteen truth valuations, which is quite distinct).  We will now show that, for any value of $j$, $\tgamma_i(F_j)=0$ when $1 \leq i \leq 12$ implies that $\tgamma_i(F_j)=0$ when $13 \leq i \leq 16$.  Thus, it cannot be true that $\tgamma_i(F_j)=1$ for at least one value of $i$ \textit{and} that $\tgamma_i(F_j)=0$ for all values of $i$ such that $\tgamma_i(A_s \cap B_s)=0$.  In other words, $F_j$ is not (non-trivially) weakly correlated to $A_s \cap B_s$.

The events $\{ F_j \}$ involve the conjunction of two indefinite events.  The truth value $\tgamma_i(P_{A0} \cap P_{A1} \cap P_{B0} \cap P_{B1})$ is the same for $i=\{4,8,12,16\}$: because of our assumptions on the truth value of conjunctions of indefinite events made above, it takes the value $w$. The truth value $\tgamma_i(P_{A0} \cap P_{A1} \cap P_{B0} \cap P_{B1})$ also holds the same value, zero, for all other values of $i$: as can be seen from the table, there are zeros in the columns for some $P$ events in all other values of $i$, which makes the conjunction definitely zero.  A similar argument can be made for all fifteen other events in the set $\{ F_j \}$.  The result is that for all values of $j$, and for $1 \leq i \leq 4$, $\tgamma_i(F_j)=\tgamma_{i+4}(F_j)=\tgamma_{i+8}(F_j)=\tgamma_{i+12}(F_j)$.  Moreover, it follows that this is true also for disjunctions of events in $\{ F_j \}$, that is, for all non-trivial past events.

From this, it follows that, for all $j$, either $\tgamma_i(F_j)=0$ for all values of 0, or it is non-zero for some $i$ in the range $1 \leq i \leq 12$.  Again, this is also true for disjunctions of events in $\{ F_j \}$.  This shows that none of the past events are (non-trivially) weakly correlated to the setting event $A_s \cap B_s$.  The same argument can be adapted to the remaining three ``full specification'' setting events $A_s \cap B_s^c$, $A_s^c \cap B_s$ and $A_s^c \cap B_s^c$.  From this, it follows that the freedom of settings condition holds.

\end{appendix}

\bibliographystyle{h-physrev3}
\bibliography{refs}

\end{document}